%% file: paper.tex
\documentclass[aps,pra, twocolumn,showpacs,floats,superscriptaddress]{revtex4-1}
\usepackage{epsfig,color}
\usepackage{bm}
\begin{document}


\author{Ping Nang Ma}
\affiliation{Theoretische Physik, ETH Zurich, 8093 Zurich, Switzerland}
\author{Lode Pollet}
\affiliation{Physics Department, Harvard University, Cambridge 02138, Massachusetts, USA}
\author{Matthias Troyer}
\affiliation{Theoretische Physik, ETH Zurich, 8093 Zurich, Switzerland}



\title{Measuring the equation of state of trapped ultracold bosonic systems in an optical lattice with in-situ density imaging}


\input{abstract}

\maketitle

\input{introduction}

\input{ultracold_bosonic_optical_lattice}

\input{fluctuation_dissipation_thermometry}
\input{second_order_high_temperature_thermometry}

\input{discussion_acknowledgement}

\input{appendix}

\end{document}

%% file: abstract.tex
\begin{abstract}

We analyze quantitatively how imaging techniques with single-site resolution allow to measure thermodynamical properties that cannot be inferred from time-of-light images for the trapped Bose-Hubbard model. If the normal state extends over a sufficiently large range, the chemical potential and the temperature can be extracted from a single shot, provided the sample is in thermodynamic equilibrium. When the normal state is too narrow, temperature is low but can still be extracted using the fluctuation-dissipation theorem over the entire trap range as long as the local density approximation remains valid, as was recently suggested by Qi Zhou and Tin-Lun Ho [arXiv:0908.3015]. However, for typical present-day experiments, the number of samples needed is of the order of 1000 in order to get the temperature at least  $10 \%$ accurate, but it is possible to reduce the variance by 2 orders of magnitude if the density-density correlation length is short, which is the case for the Bose-Hubbard model. Our results provide further evidence that cold gases in an optical lattices can be viewed as quantum analog computers.

\end{abstract}

%% file: introduction.tex
\section{Introduction}
The prototypical, unsolvable models of condensed matter physics can be implemented in a tunable, clean and controllable way with cold atoms in optical lattices~\cite{review}. Experiments have the potential to offer new insight in the long-standing problems of condensed matter physics, and would ultimately allow us to establish the validity of those models in the description of more complicated materials. Before such quantum simulators can be trusted they need to be validated by benchmarking them against known results of models that can be solved accurately on a classical computer. This was done in great detail for the superfluid to normal liquid transition at unity filling in the Bose-Hubbard model by comparing experimentally observed interference patterns to the ones computed in full ab-initio quantum Monte Carlo simulations~\cite{Trotzky09}.

One important experimental issue is the accurate determination of temperature. A thermometer measures a property of a subsystem in thermal equilibrium with the rest of the system. The measurement of this property can be used to determine the temperature if the temperature dependence of this property is known from theoretical calculations or experimental calibration. The accuracy of the thermometer increases and the number of measurements necessary decreases if more information about the system is available.

In this context, recent experimental progress in single-site resolution and addressability~\cite{Chin2009, Greiner2009, Wuertz2009, Bakr2010, Sherson2010} provides us with important additional tools to validate the Bose-Hubbard model. With this it is possible to determine temperature in the lattice system using the fluctuation-dissipation theorem without invoking theory~\cite{Zhou09,Sanner, Muller}. Also the equation of state and the chemical potential can be extracted from the edges using high temperature series expansions, provided those regions are large enough. This was previously suggested when the edges are in the ideal gas regime~\cite{Zhou_Kato, Ho_Zhou}; however this is too restrictive. We note that number fluctuations  were previously suggested as an effective thermometer~\cite{Capogrosso07, Gerbier06}. A failure to extract temperature and/or chemical potential signals that the experiment is not in equilibrium. 


In this paper, we extend our ab-initio study of ultra-cold gases~\cite{Trotzky09} to the physics accessible through single-site resolution detection tools. We focus on extracting the temperature from the lattice experiments. In doing so, we extend the studies initiated in Refs.~\cite{Zhou_Kato, Ho_Zhou} from the ideal gas to the strongly interacting systems for realistic system sizes, and hereby validate and improve on their proposals where applicable. Carefully choosing the range over which correlations are measured we show that accurate thermometry can be performed with a few dozen measurements. Using additional theoretical input, such as a second order high temperature expansion of the density, it becomes possible - in some parameter regimes - to extract the temperature accuractely from a single density profile.

The structure of this paper is as follows.
First, we introduce the model in the next section.
Second, we discuss a general thermometry scheme based on the fluctuation-dissipation (FD) theorem, and illustrate its validity for strongly correlated bosons in an optical lattice. Third, we investigate how reliably $2^{nd}$ order high temperature series expansions (HTE2) can be applied to the edges in order to extract the temperature and the chemical potential, before concluding in Section IV.

%% file: ultracold_bosonic_optical_lattice.tex
\section{Ultracold bosonic optical lattice}

Strongly interacting bosons harmonically trapped in an isotropic optical lattice can be quantitatively described by the single-band  boson Hubbard Hamiltonian~\cite{Jaksch}, 
\begin{equation}
\hat{H} - \mu \hat{N} = - t\sum_{\langle i,j \rangle} \hat{b}_i^\dag \hat{b}_j + \frac{U}{2} \sum_i \hat{n}_i (\hat{n}_i - 1) - \sum_i (\mu - V_T r_i^2) \hat{n}_i,
\end{equation}
where $\hat{b}$ ($\hat{b}^\dag$) and $\hat{n}$ are annihilation (creation) and number operators, respectively. The operator $\hat{N}$ counts the total number of particles found for a system with chemical potential $\mu$, and the external harmonic trapping is characterized by $V_T$. The nearest neighbor hopping $t$,  the onsite repulsion strength $U$, and the confinement strength $V_T$ are derived from the lattice laser potential $V_0$ through band structure calculations~\cite{Jaksch}.

To mimic in-situ density measurements in ultracold bosonic optical lattice experiments, we have performed Quantum Monte Carlo worm simulations\cite{Nikolay,Lode} (exact up to the statistical errorbars) and obtained a time-series of 3-dimensional density measurements  which are then column-integrated along the line-of-sight direction.
Uncorrelated measurements are obtained by imposing a strict criterion for the autocorrelation time $\tau < 0.2$ on our simulation data used in subsequent thermometry analysis.

Throughout this paper, we focus on a physical system of 125,000 ${}^{87}$Rb atoms. We choose realistic parameters, using a trapping frequency $V_T/t=0.0091$ for $U/t=10$ and $V_T/t=0.0277$ for $U/t=50$.
For convenience, we use the lattice spacing $a = \lambda/2$ as unit of length.

%% file: fluctuation_dissipation_thermometry.tex
\section{Thermometry scheme based on the fluctuation-dissipation theorem}
\label{sect:FD}

\begin{figure}
\centering
\includegraphics[width=4cm]{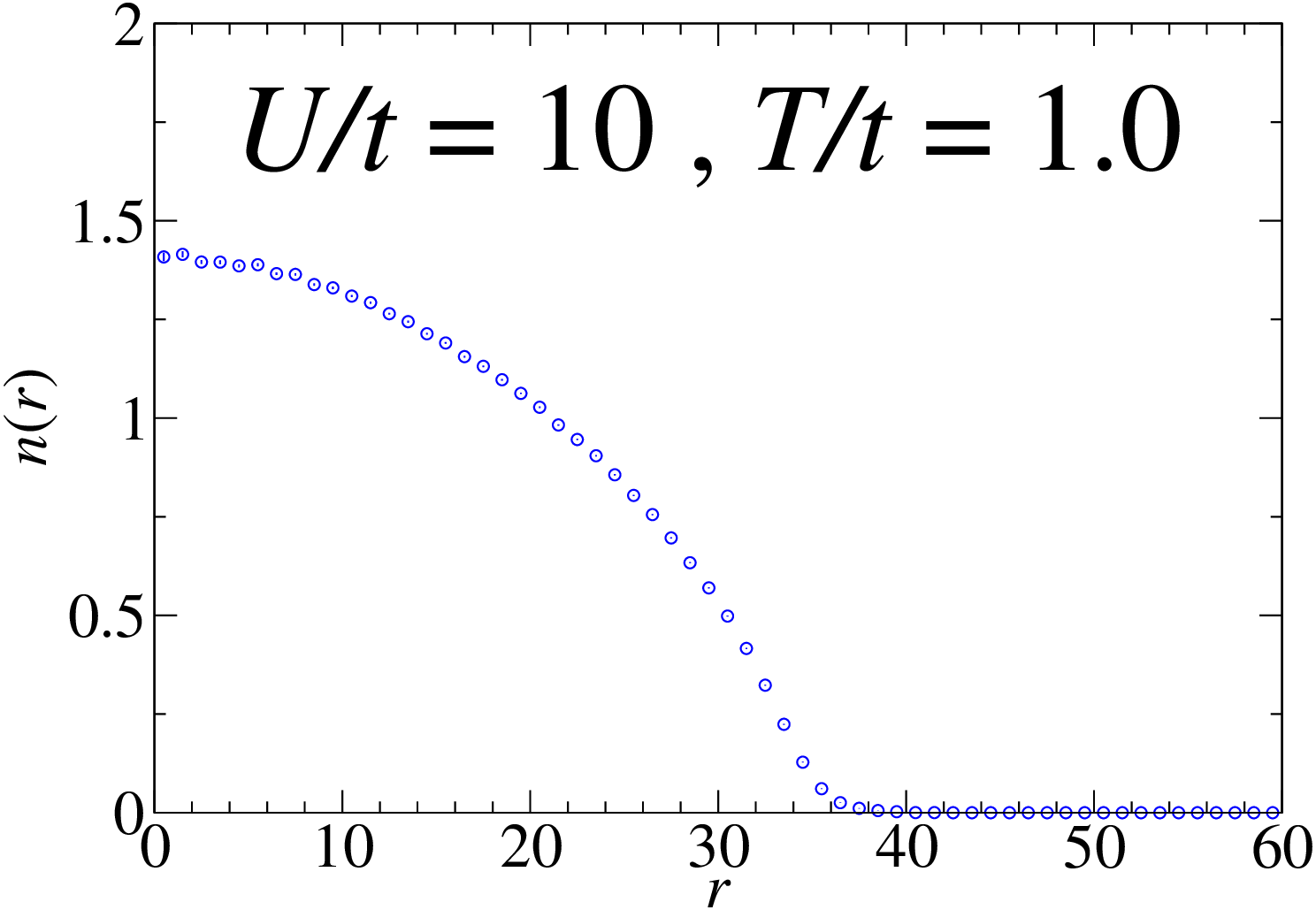}
\hfill
\includegraphics[width=4cm]{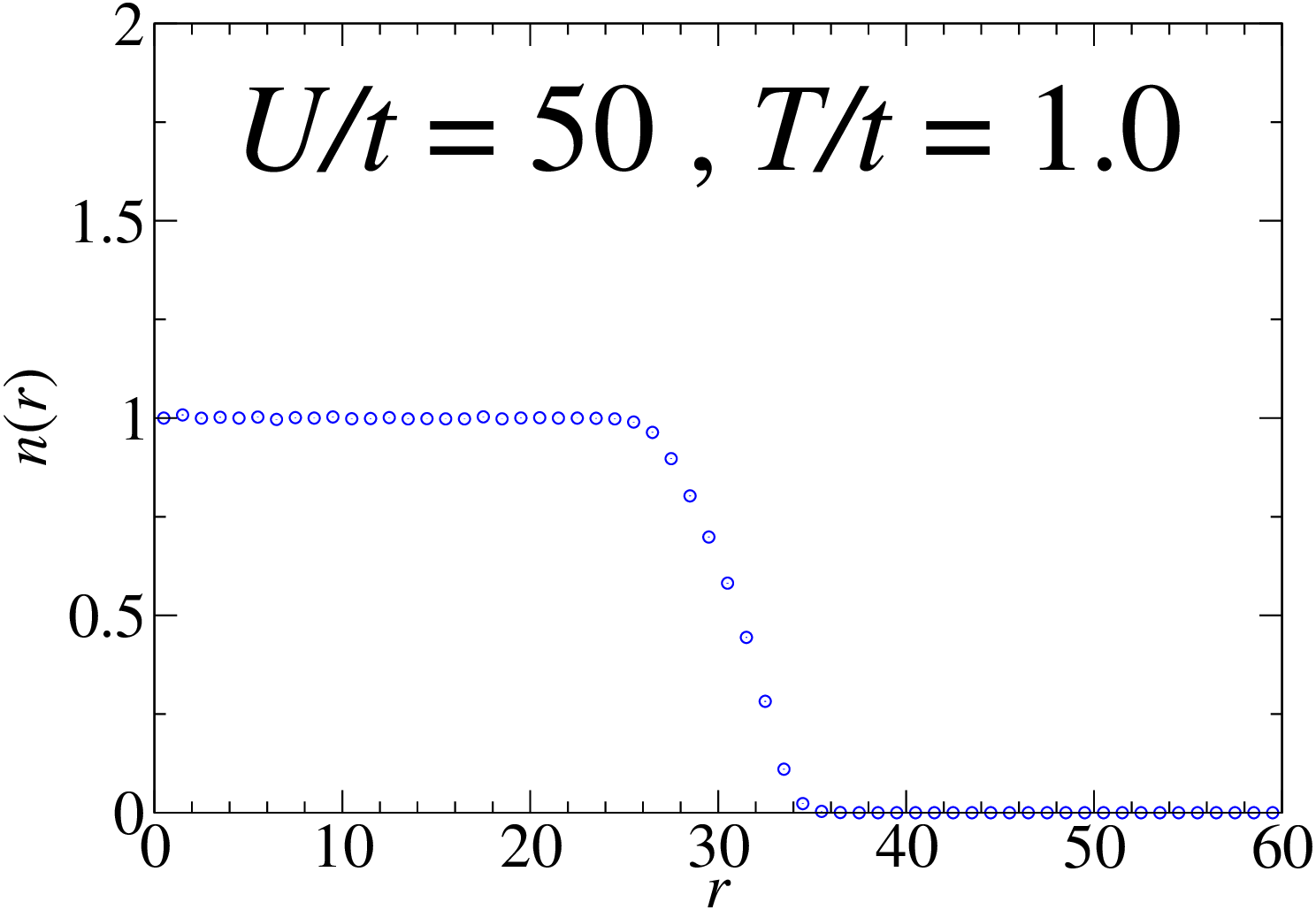}
\hfill
\includegraphics[width=4cm]{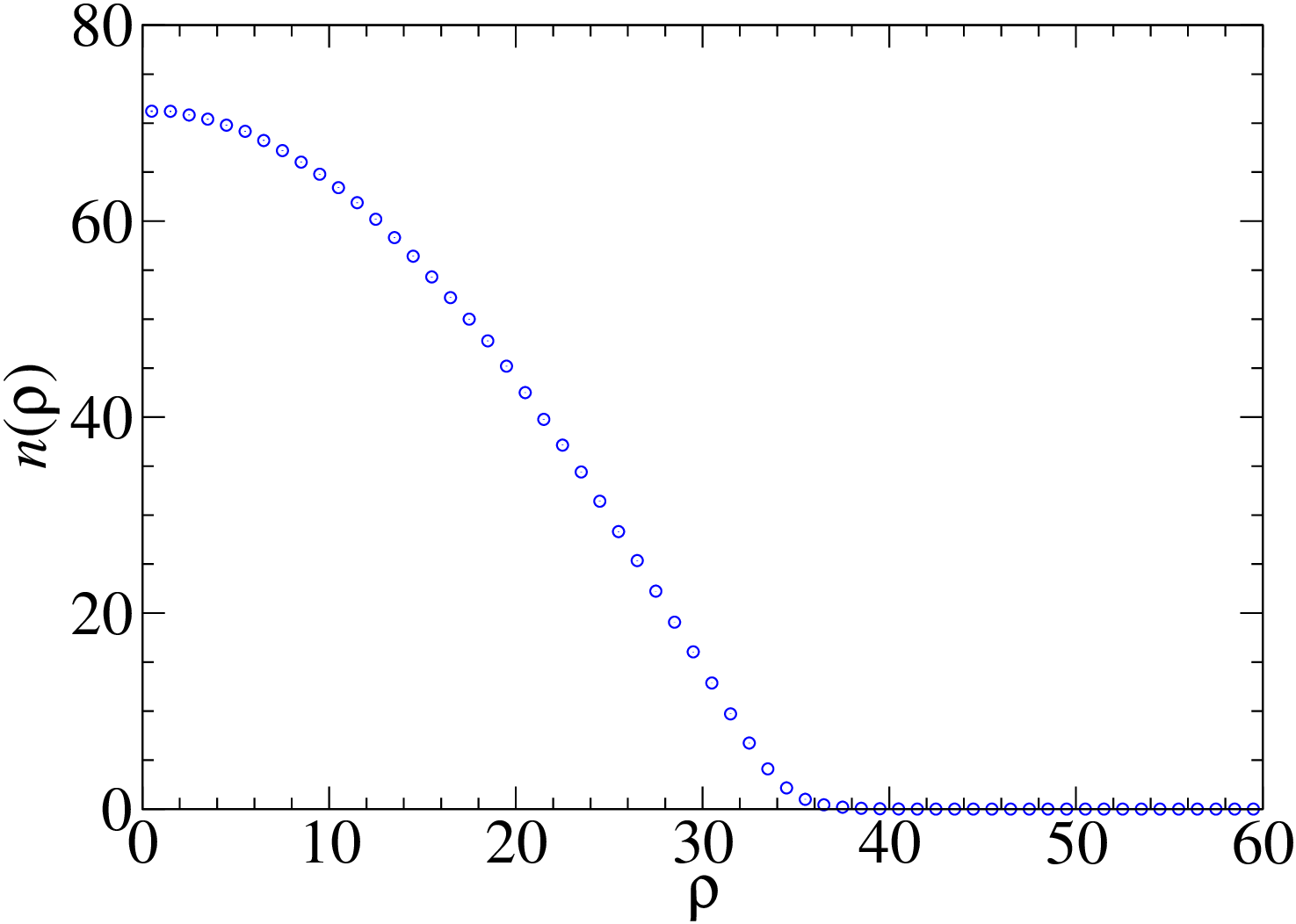}
\hfill
\includegraphics[width=4cm]{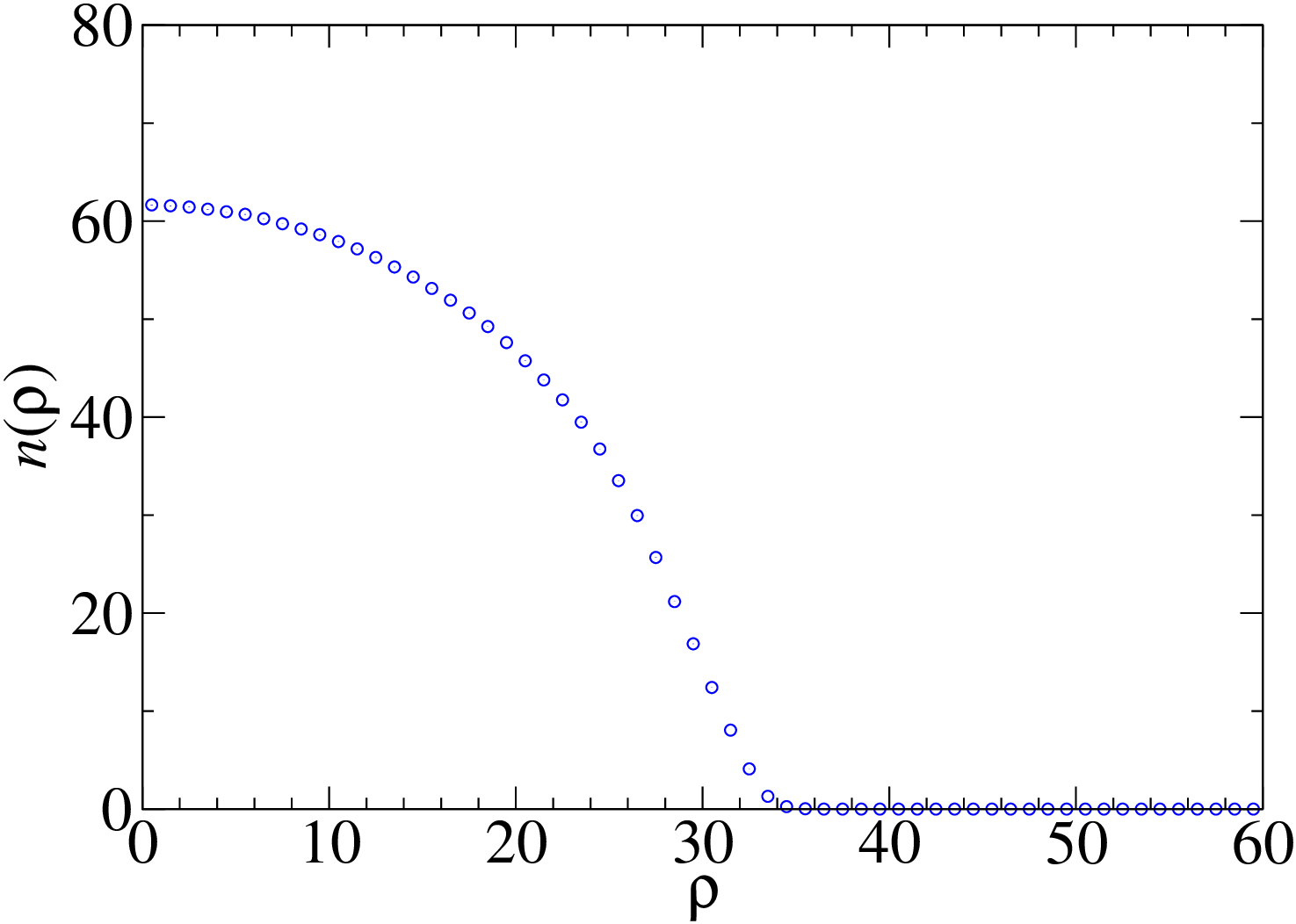}
\hfill
\includegraphics[width=4cm]{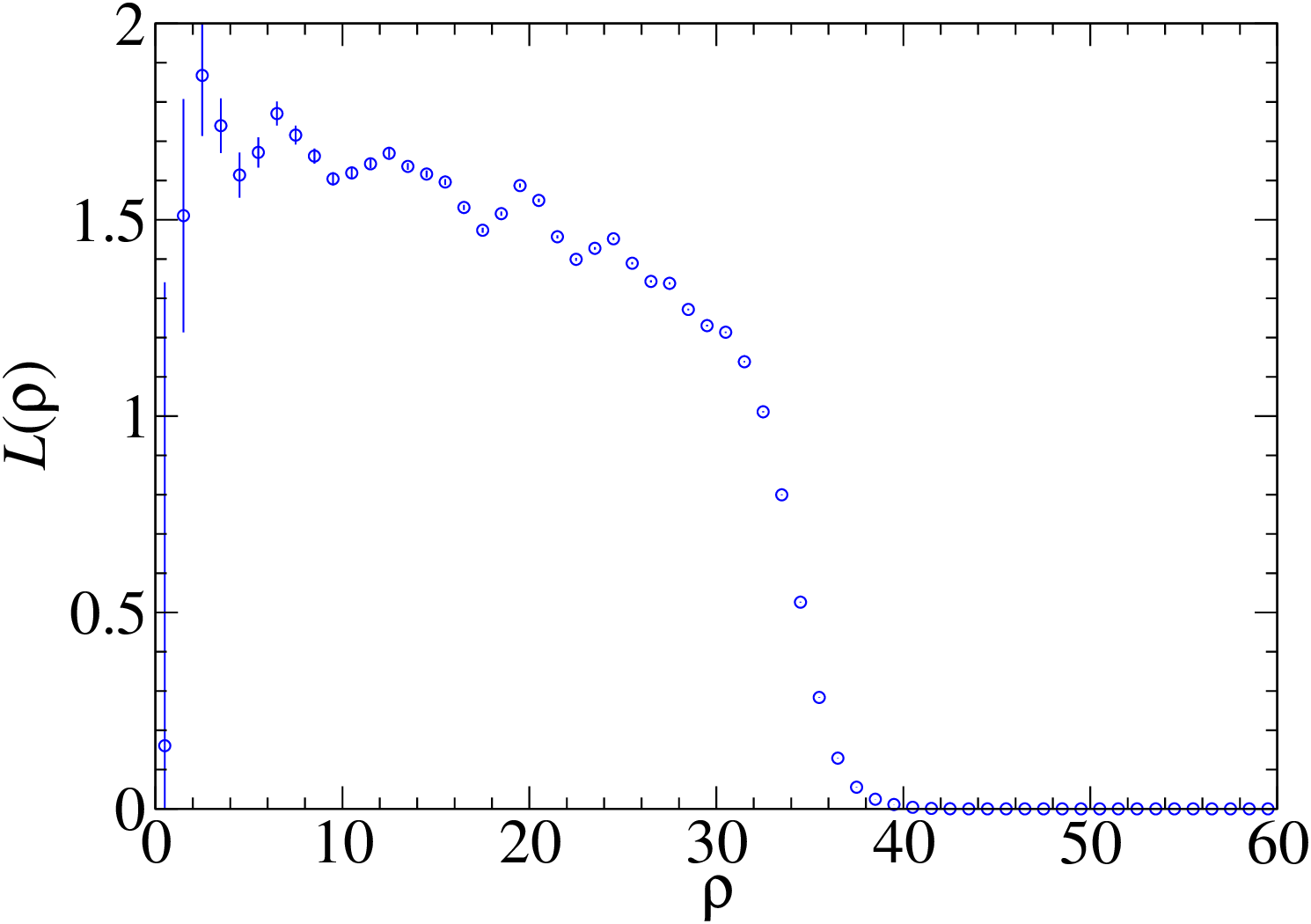}
\hfill
\includegraphics[width=4cm]{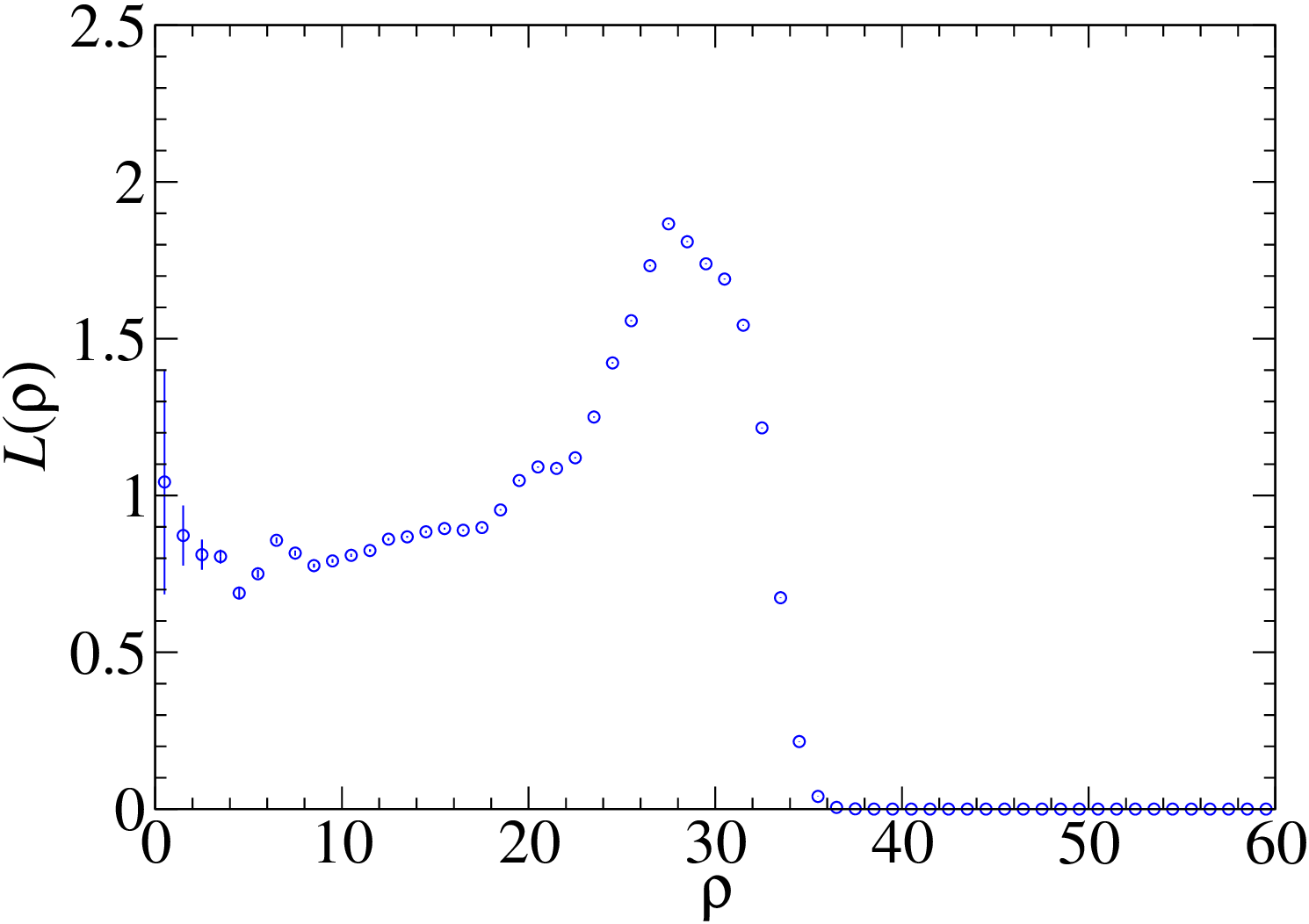}
\hfill
\includegraphics[width=4cm]{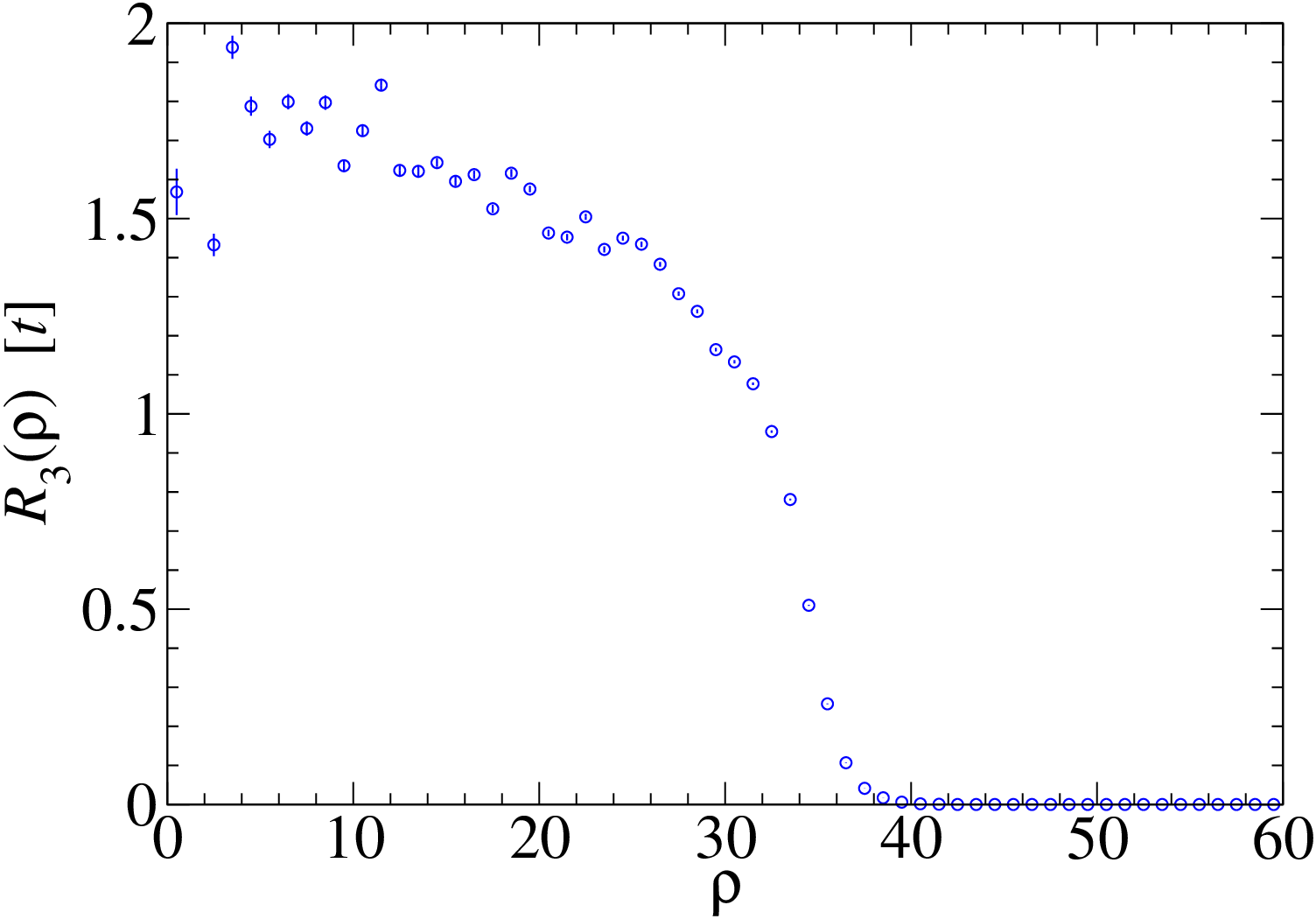}
\hfill
\includegraphics[width=4cm]{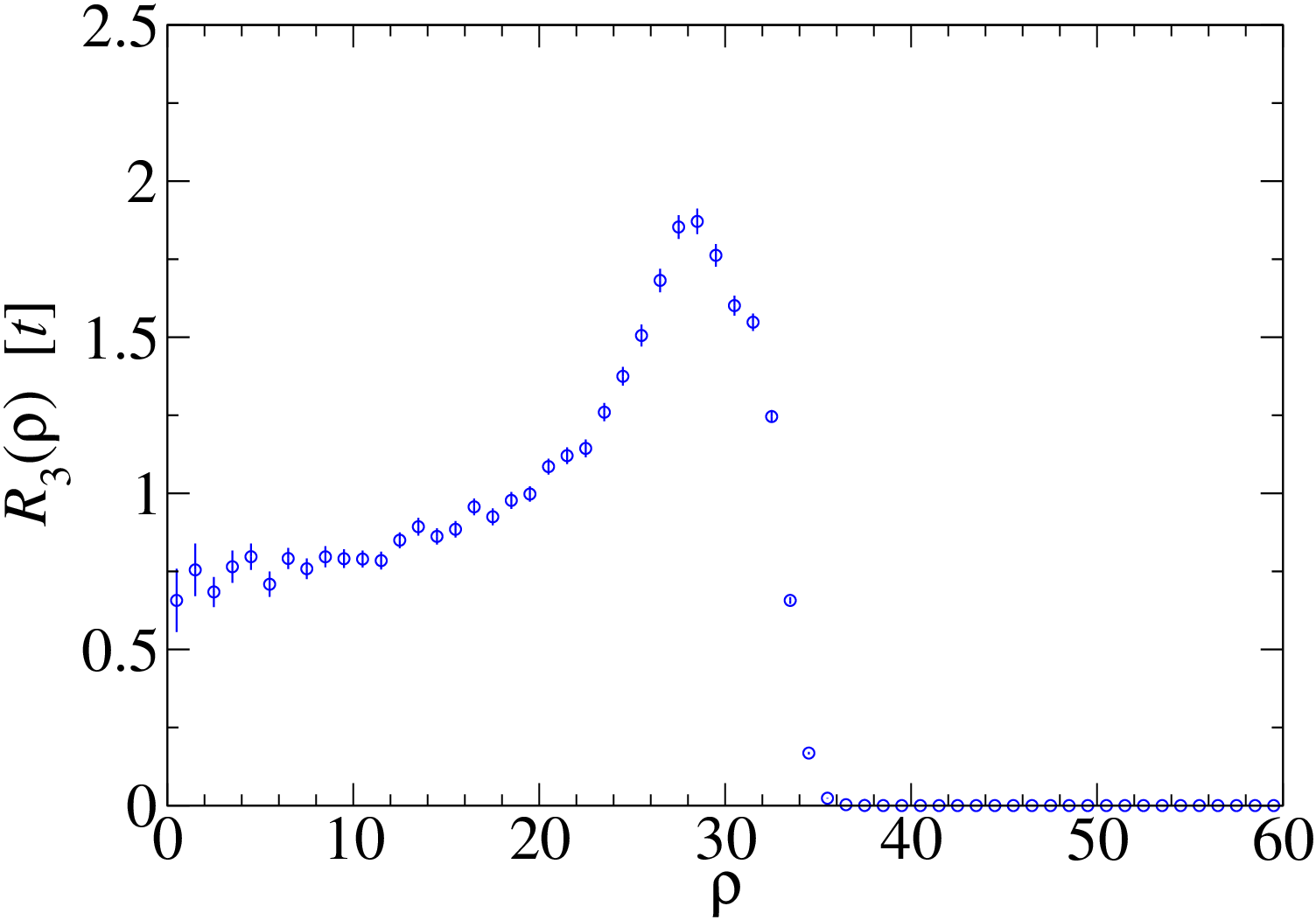}
\hfill
\includegraphics[width=4cm]{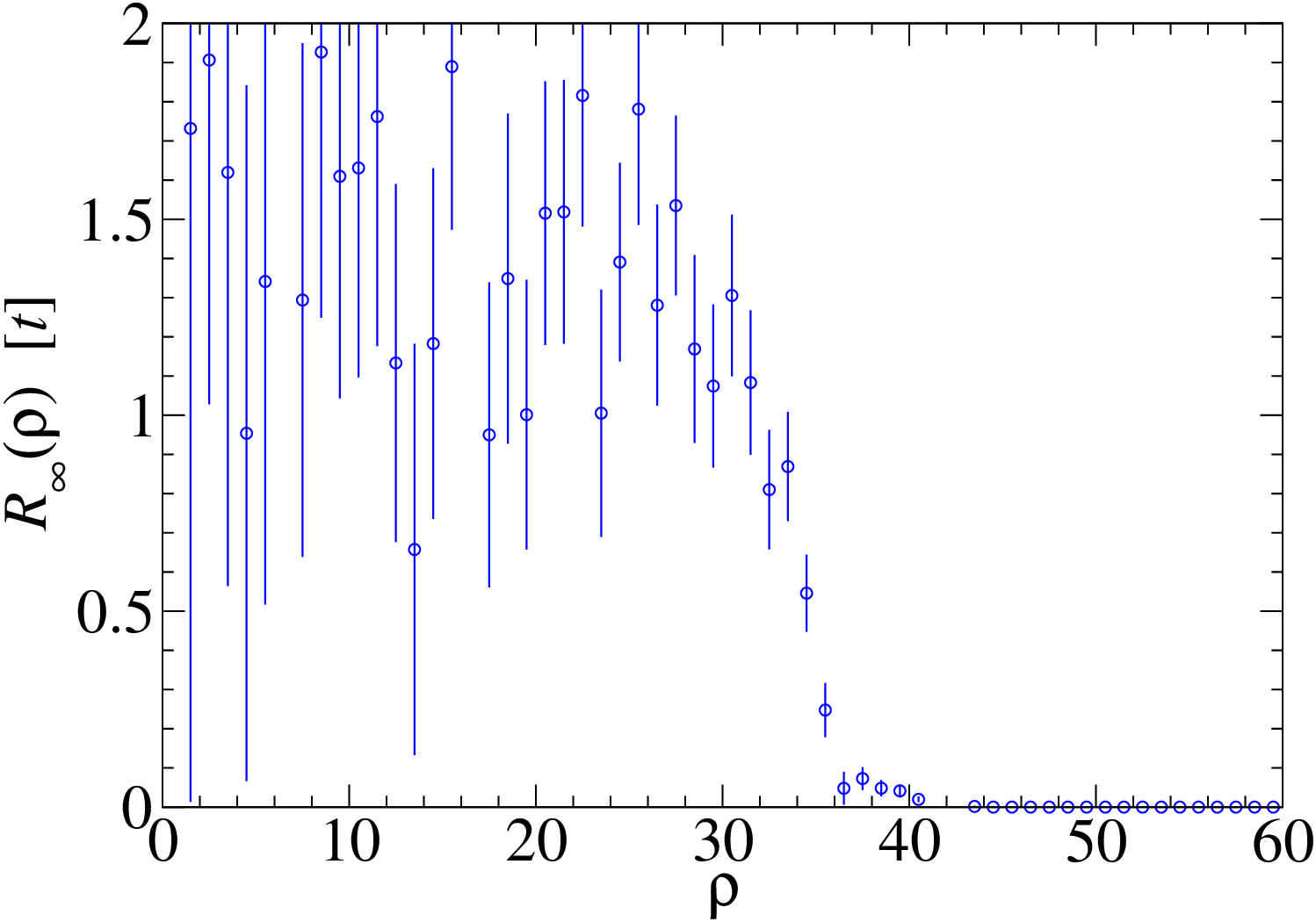}
\hfill
\includegraphics[width=4cm]{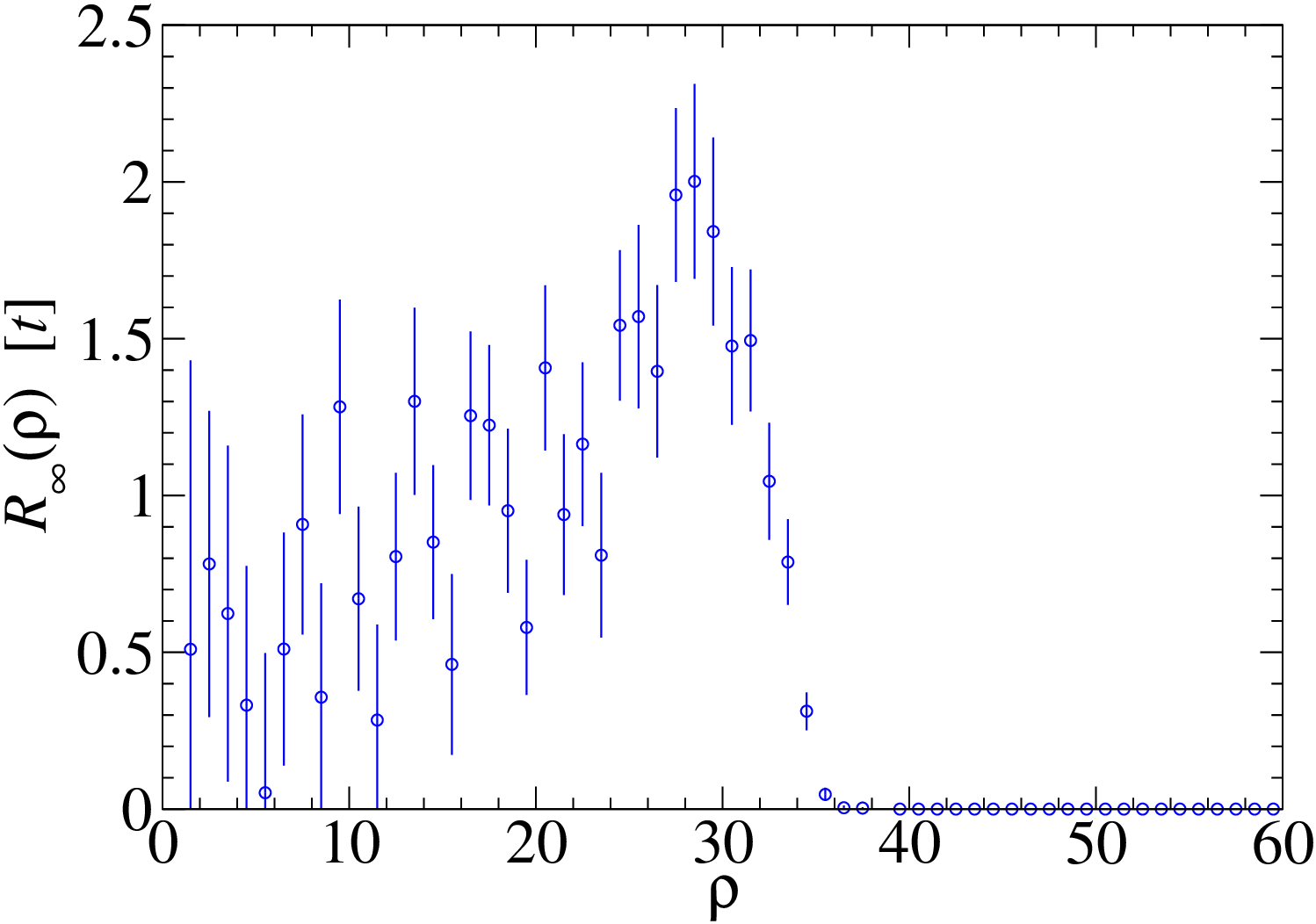}
\caption
{
Illustration of the quantities entering the fluctuation-dissipation thermometry formula. Shown from top to bottom are : 1) cross-section density $n(r)$ , 2) column integrated density $n(\rho)$ , 3) dissipation term $L(\rho)$ , 4) fluctuation term ($\xi=3$) $R_3(\rho)$ , 5) fluctuation term ($\xi=\infty$) $R_\infty(\rho)$.
We take a 3D bosonic ${}^{87}$Rb optical lattice system with $N = 125,000$ and we average over 1000 independent measurements obtained from a QMC simulation.
The parameters in the left column are $U/t = 10$ , $T/t = 1$; and in the right column: $U/t = 50$ , $T/t = 1$.
}
\label{fig:overall_FD}
\end{figure}

\begin{figure}
\centering
\includegraphics[width=4cm,height=2.8cm]{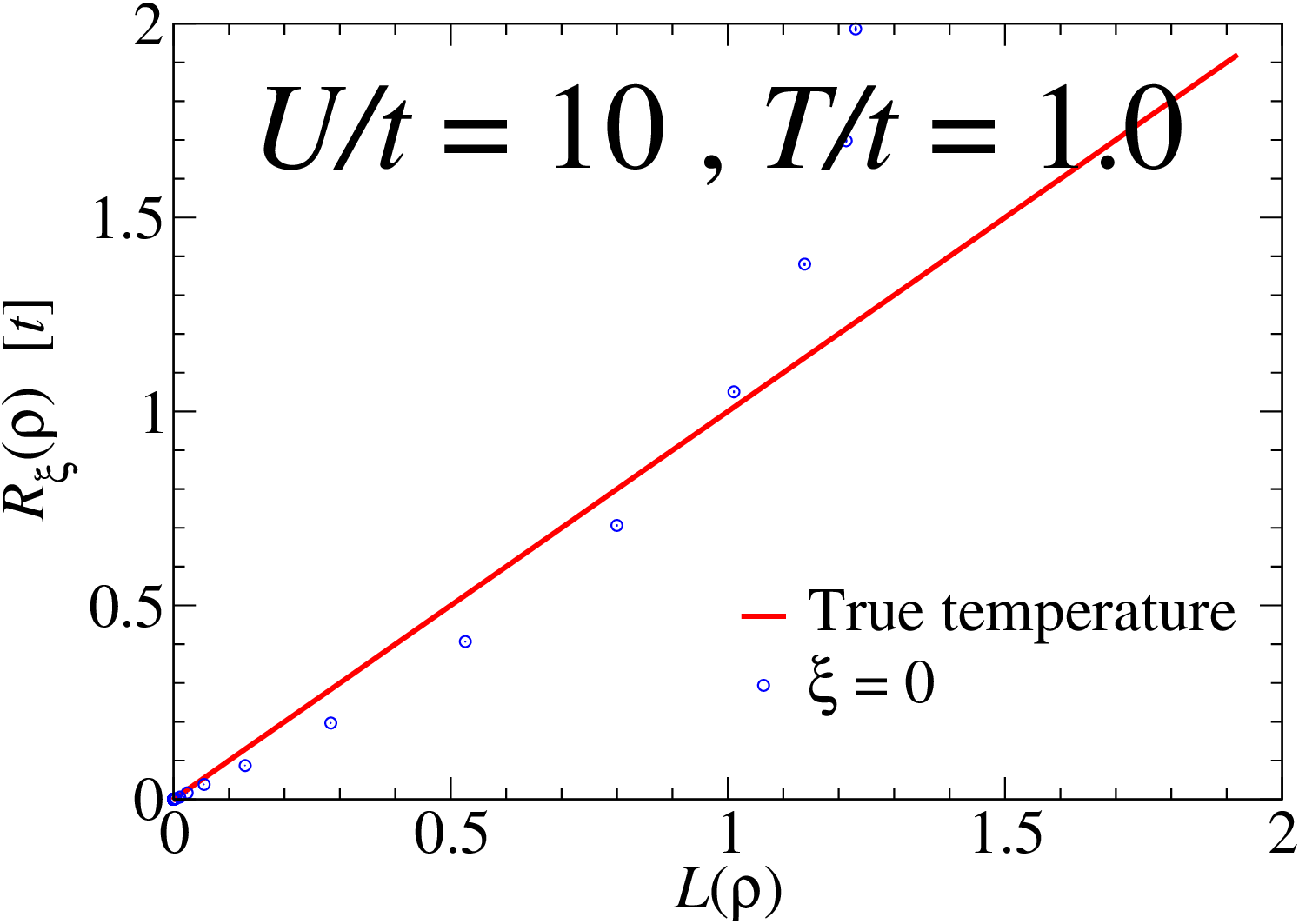}
\hfill
\includegraphics[width=4cm,height=2.8cm]{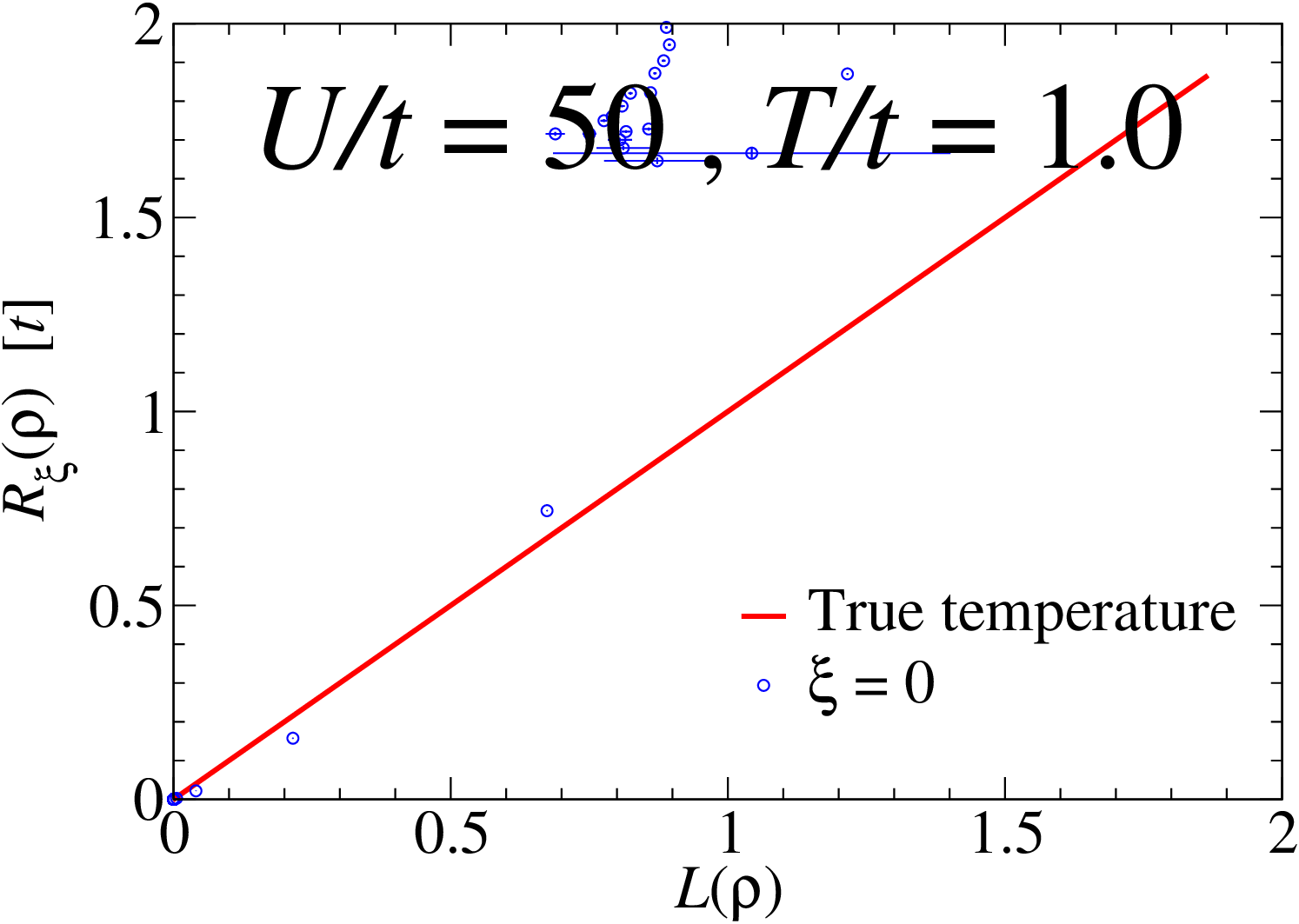}
\hfill
\includegraphics[width=4cm,height=2.8cm]{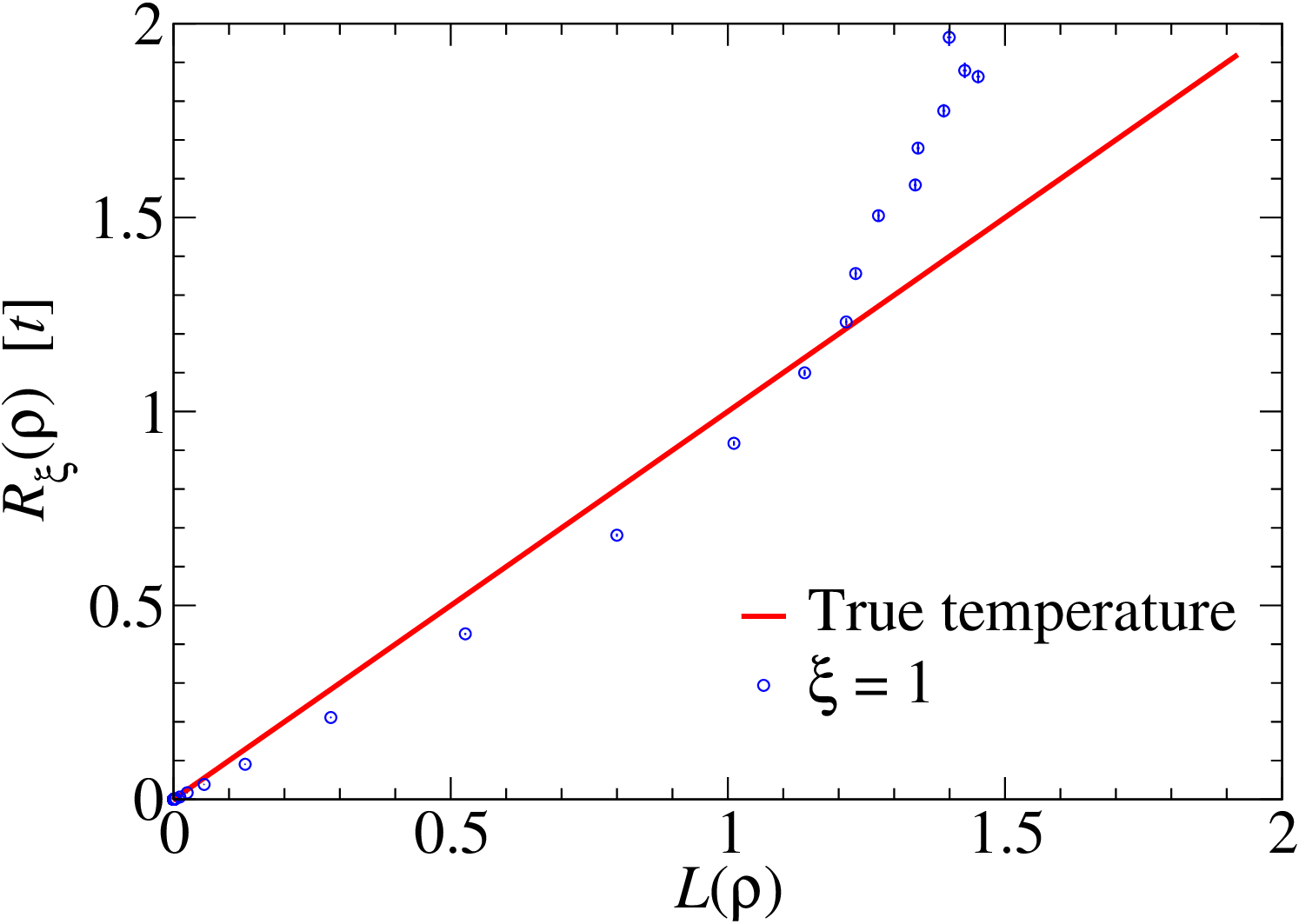}
\hfill
\includegraphics[width=4cm,height=2.8cm]{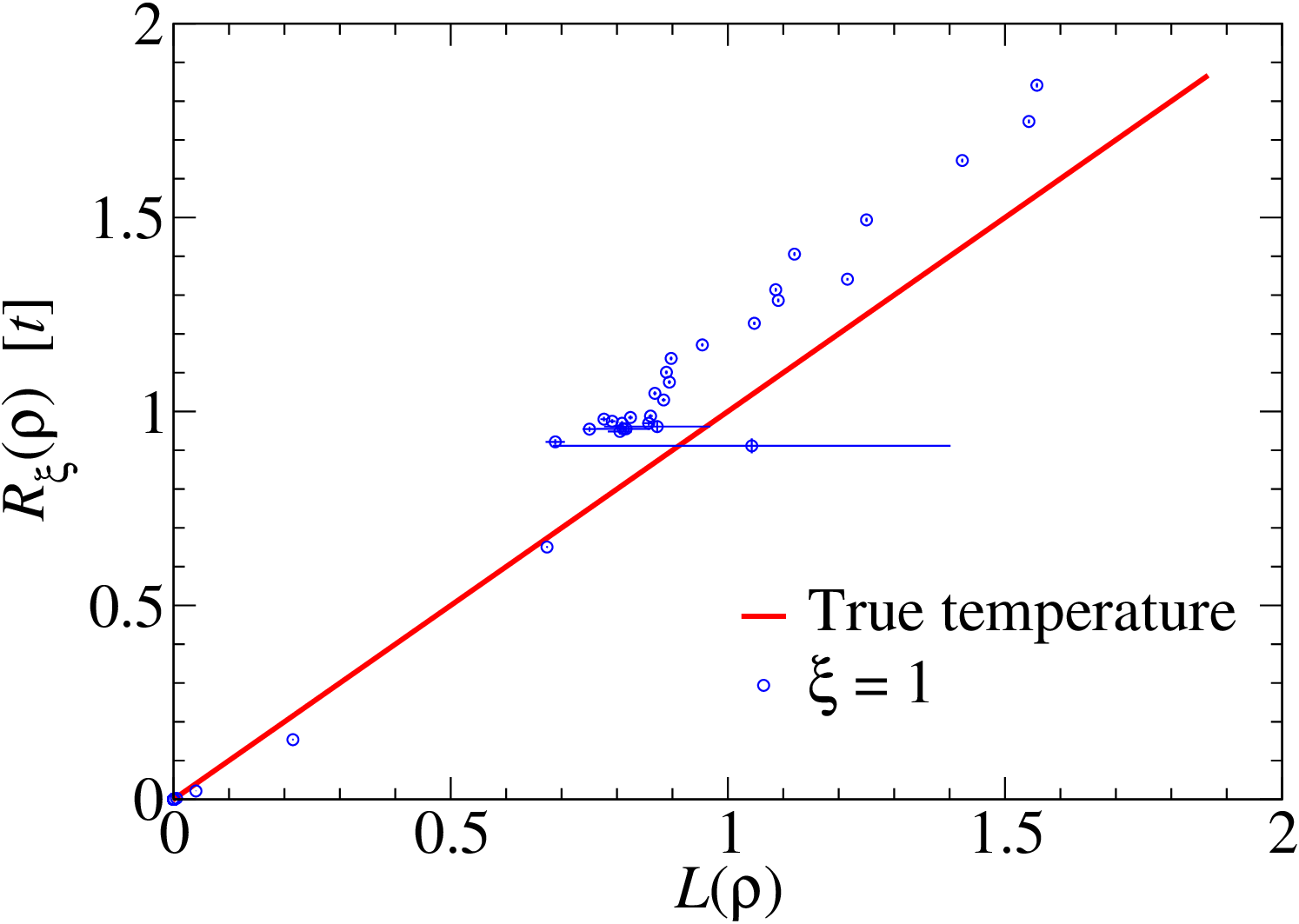}
\hfill
\includegraphics[width=4cm,height=2.8cm]{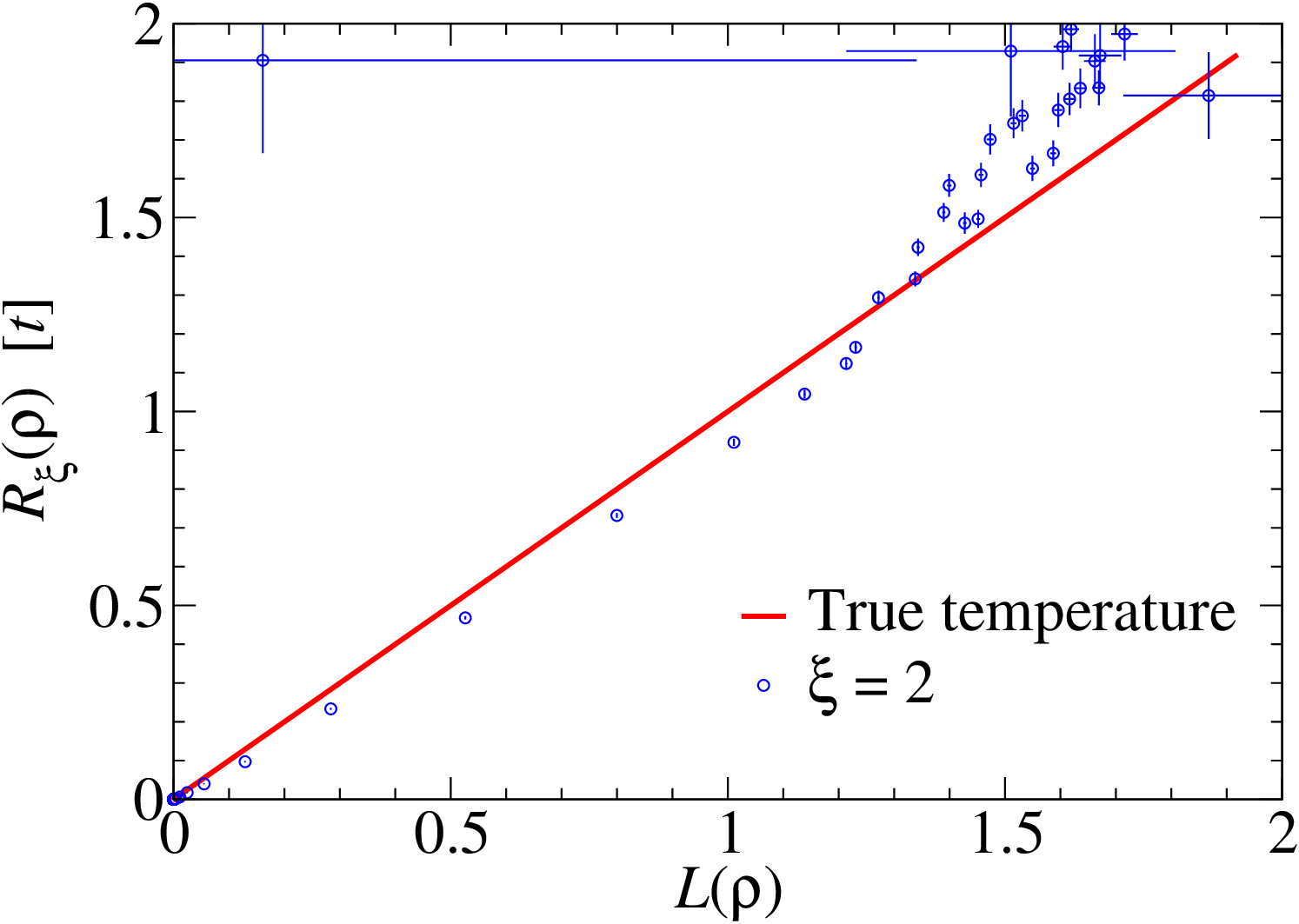}
\hfill
\includegraphics[width=4cm,height=2.8cm]{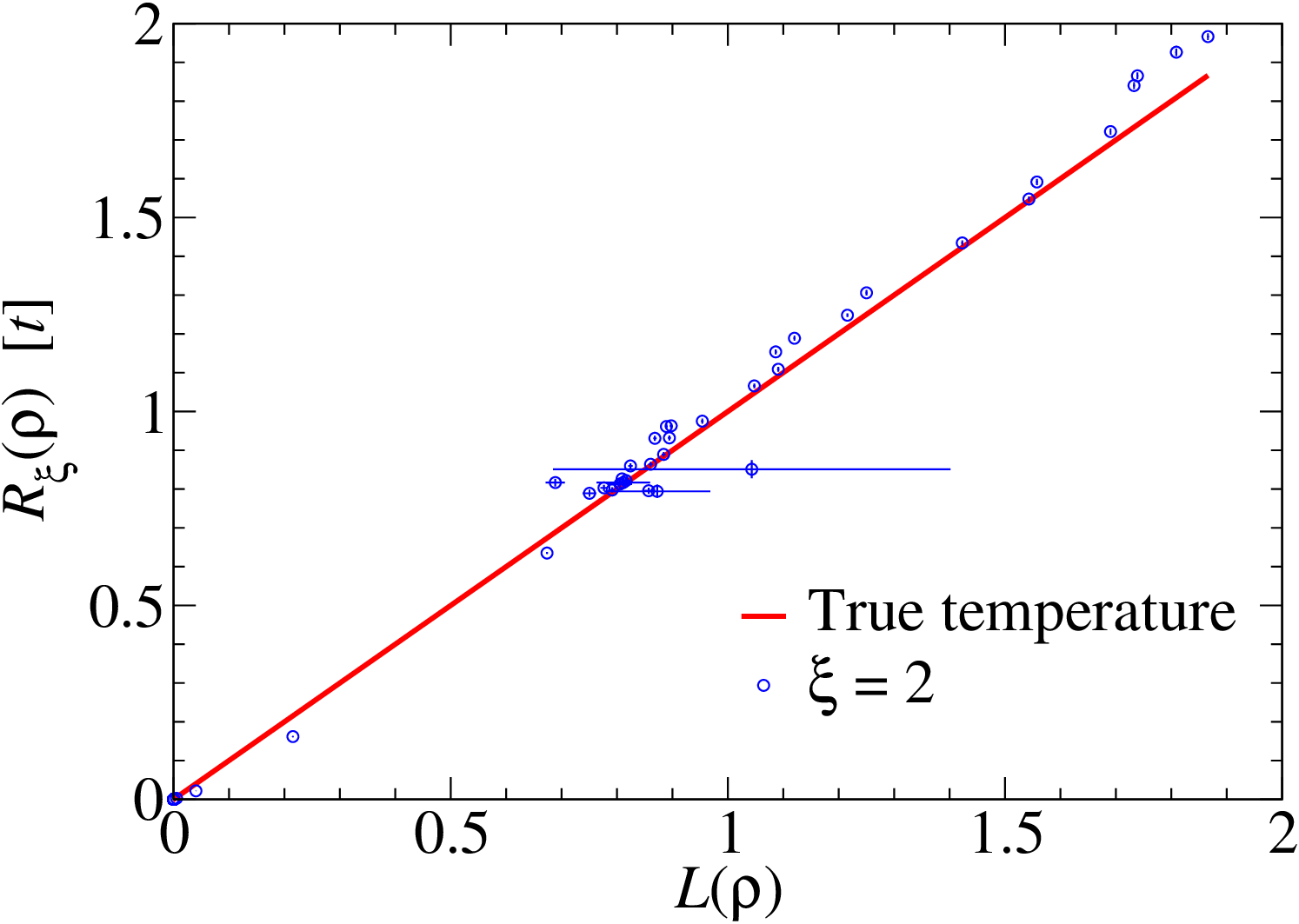}
\hfill
\includegraphics[width=4cm,height=2.8cm]{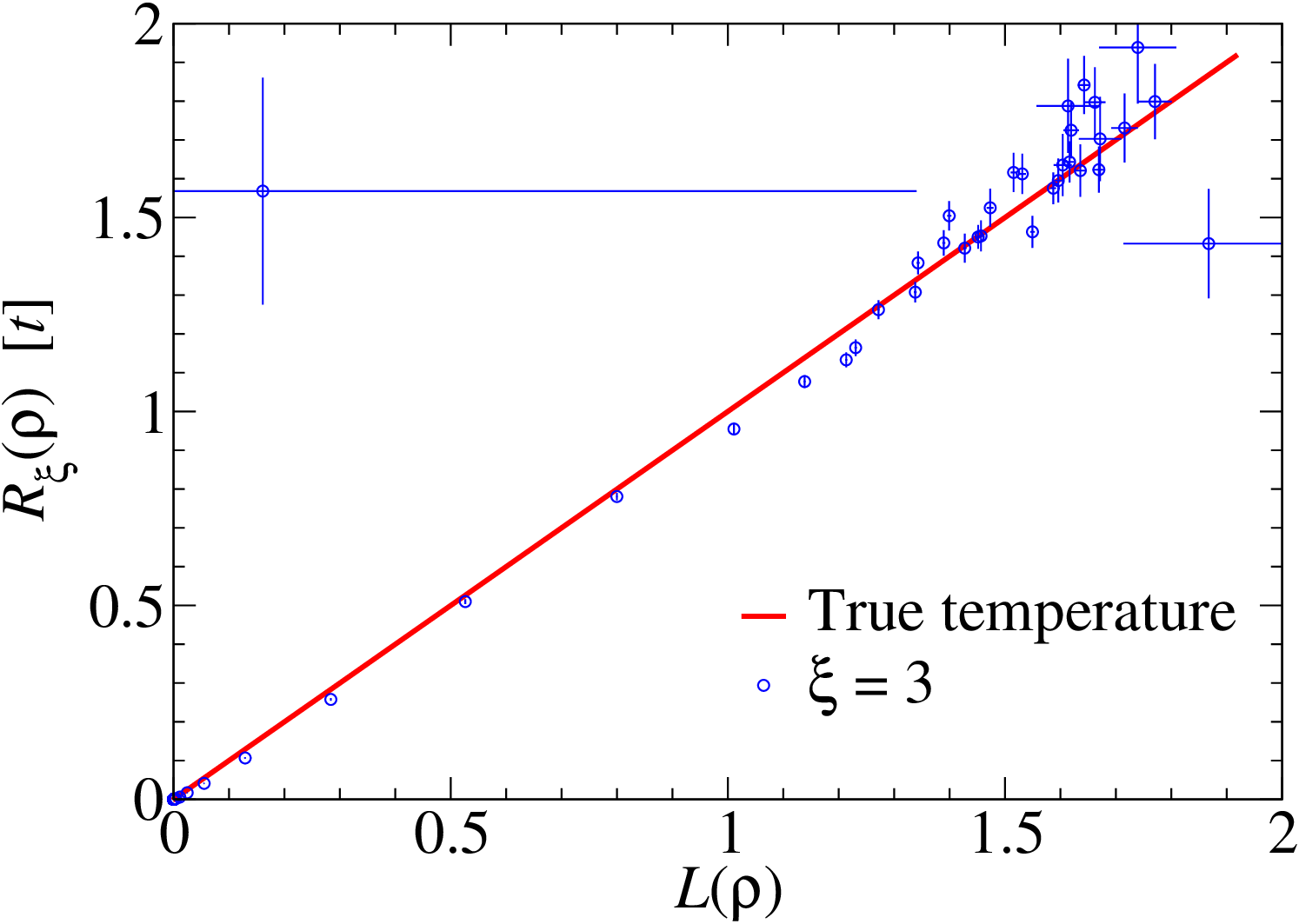}
\hfill
\includegraphics[width=4cm,height=2.8cm]{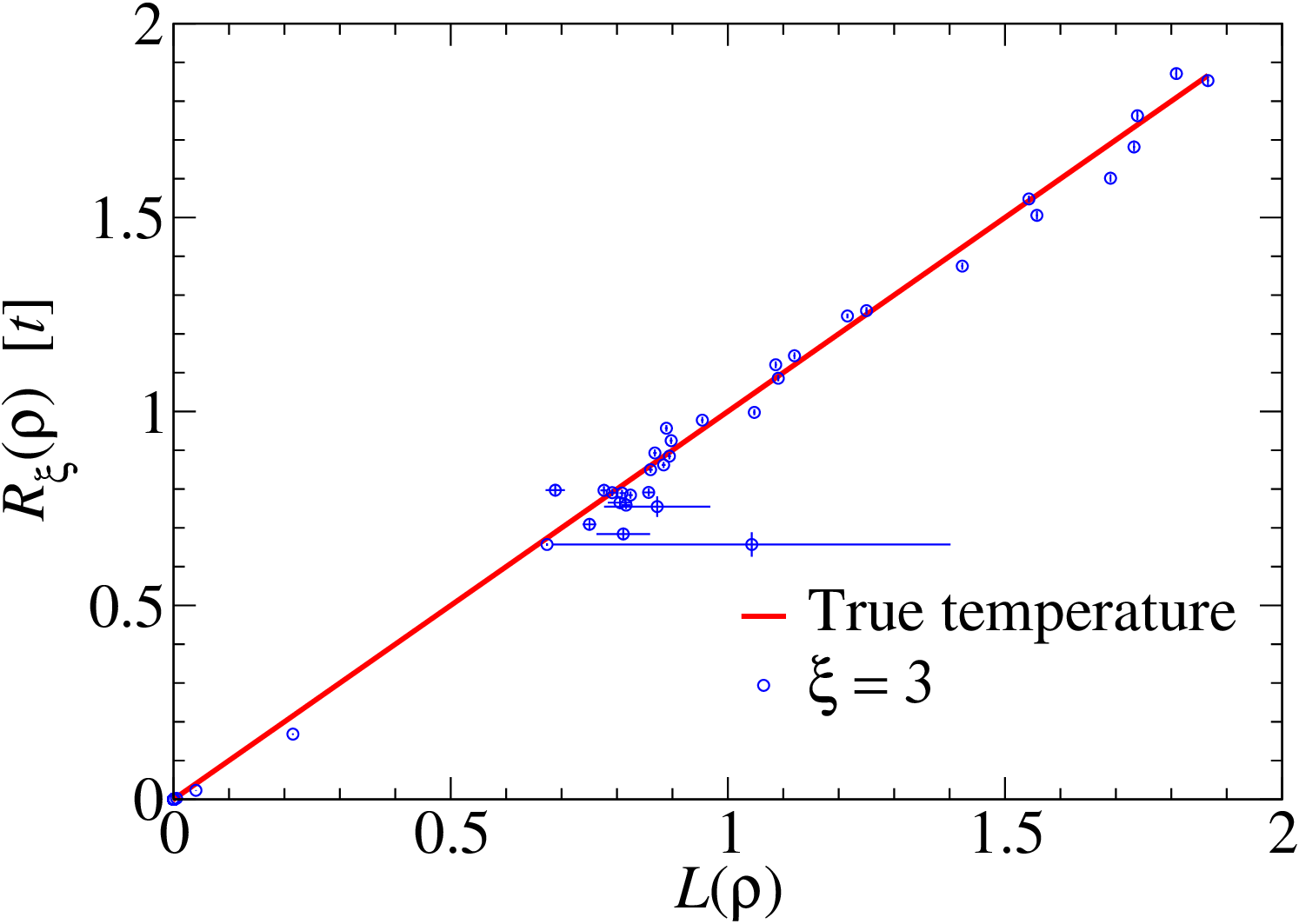}
\hfill
\includegraphics[width=4cm,height=2.8cm]{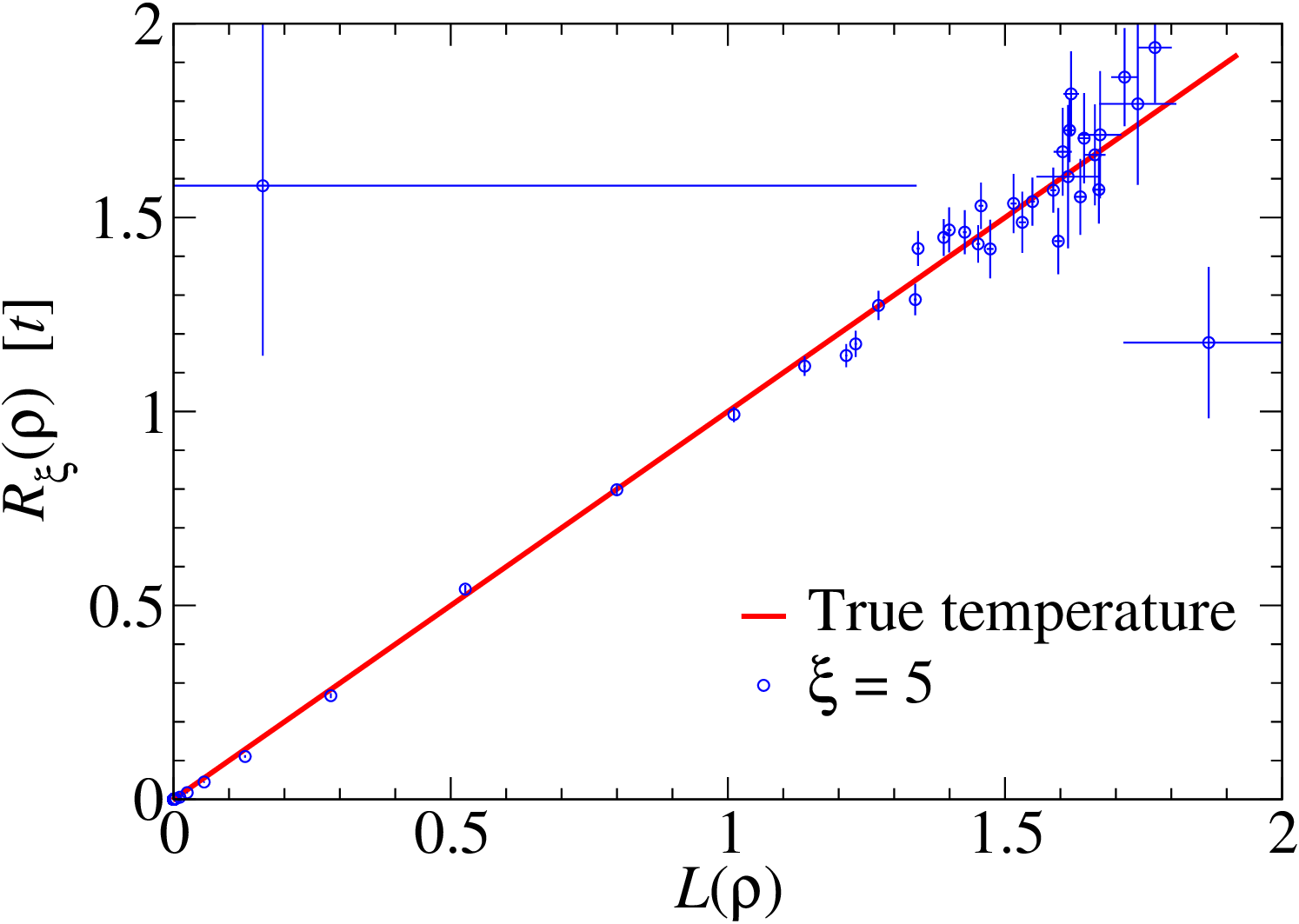}
\hfill
\includegraphics[width=4cm,height=2.8cm]{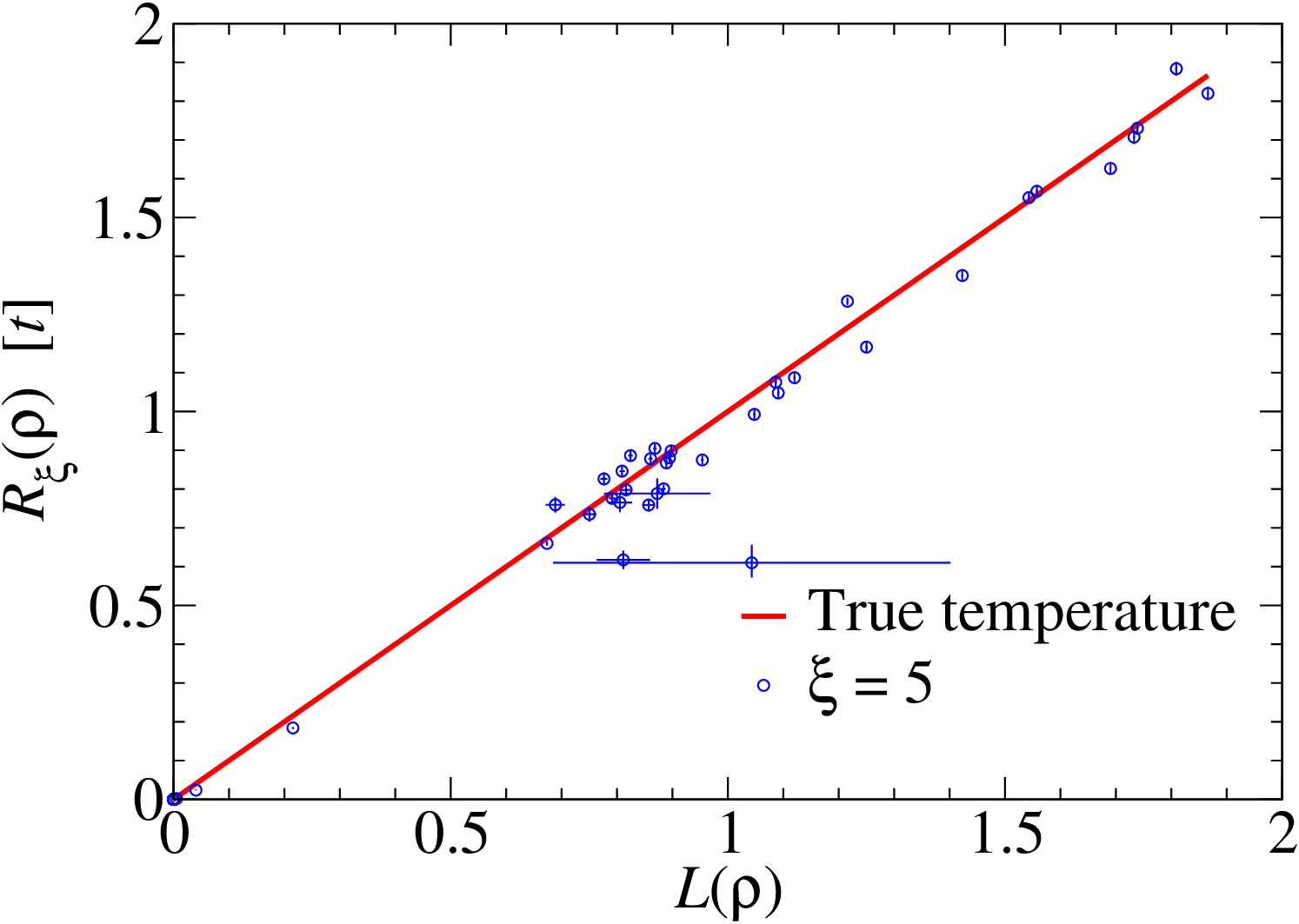}
\hfill
\includegraphics[width=4cm,height=2.8cm]{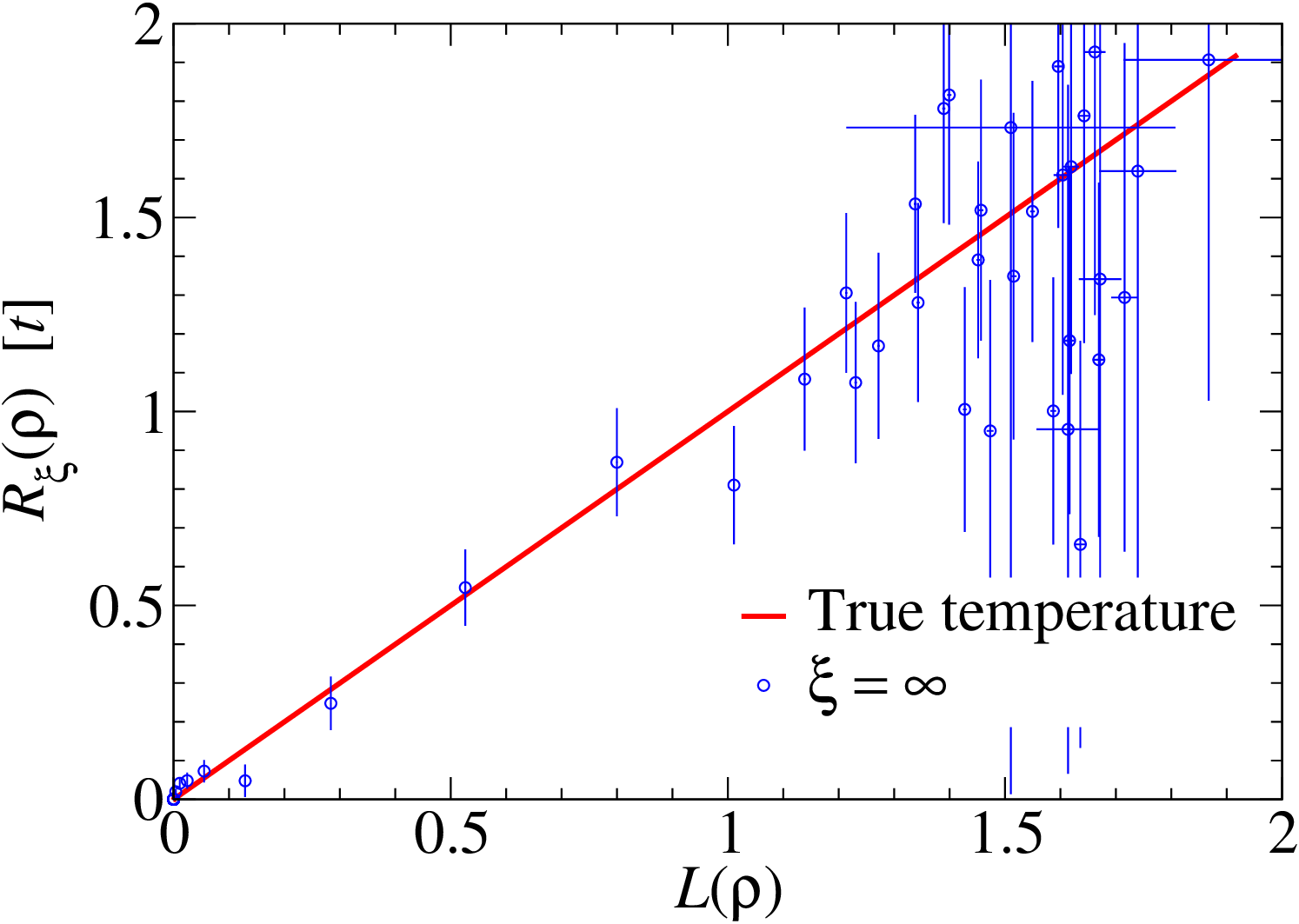}
\hfill
\includegraphics[width=4cm,height=2.8cm]{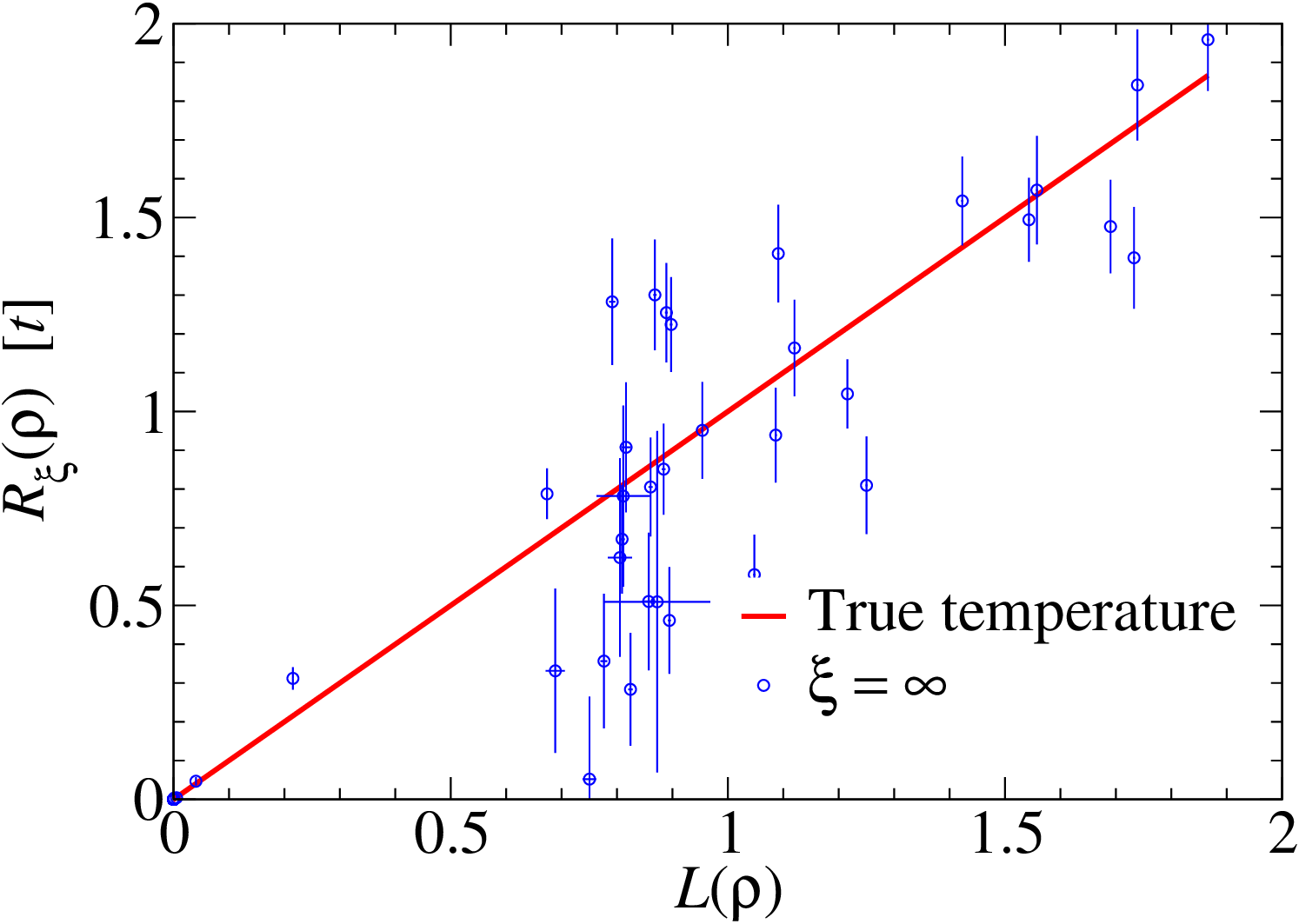}
\caption
{
Illustration of the fluctuation-dissipation thermometry scheme by showing different $L(\rho)$-$R_\xi(\rho)$ plots at various window sizes  $\xi=0,1,2,3,5,\infty$.
This approach illustrates how the density-density correlation length can be found in an experimental system.
When the window size is smaller than density-density correlation length, systematic errors set in and this results in nonlinear $L(\rho)$-$R(\rho)$ behavior, while for $\xi$ larger than the density-density correlation length the behavior of $L(\rho)-R(\rho)$ is linear. However, statistical noise also increases with increasing window size.
We take a 3D bosonic ${}^{87}$Rb optical lattice system with $N = 125,000$ and we average over 1000 independent measurements obtained from a QMC simulation.
The parameters in the left column are $U/t = 10$, $T/t = 1$;
 and in the right column $U/t = 50$, $T/t = 1$.
}
\label{fig:03}
\end{figure}

In this section, we illustrate how the fluctuation-dissipation (FD) theorem can be turned into an effective temperature probe for ultracold bosonic optical lattices, enabled by in-situ density imaging experiments~\cite{Chin2009, Greiner2009}. Our approach is a generalization of the proposal of Zhou and Ho\cite{Zhou09}. It is based upon the 3D density,  
\begin{equation}
\label{eqdns}
 \langle n ({\bf r}) \rangle = \langle n (\rho, \phi, z; T, \mu) \rangle = \frac{{\rm Tr} \,  \hat{n}(\rho, \phi, z) \, e^{-\beta(\hat{H} -\mu \hat{N}) } }  {{\rm Tr} \, e^{-\beta(\hat{H} -\mu {\hat{N}}) }},
\end{equation}
integrated along the line-of-sight $\langle n ({\bm \rho}) \rangle = \int dz \, \langle n ({\bf r}) \rangle$, and the integrated density-density correlations,  
\begin{equation}
\label{fluct_eq}
R_\xi({\bm \rho}) = \int d{\bm \rho}' \, \{ \, \langle n({\bm \rho}) n({\bm \rho}') \rangle - \langle n({\bm \rho}) \rangle \langle n({\bm \rho'}) \rangle \, \} \, \theta \left( \xi - \left| {\bm \rho} - {\bm \rho}' \right| \right),
\end{equation}
within a window size $\xi$. Both quantities can be measured directly from in-situ density images. Here, $\mathbf{r}$ denotes a coordinate in three dimensions parametrized in cylindrical coordinates as $\mathbf{r}(\rho, \phi, z)$, while ${\bm \rho}(\rho, \phi)$ denotes the in-plane coordinate. $\theta(.)$ is the Heaviside step function. 
Under the assumptions of the validity of the local density approximation (LDA) and a value of $\xi$ that is larger than the density-density correlation length, the FD theorem takes the form  $T \times L({\bm \rho}) = R_\xi({\bm \rho})$ where the dissipation term
\begin{equation}
\label{diss_eq}
L({\bm \rho}) = \left(\frac{\partial \langle n( {\bm \rho}) \rangle}{\partial \mu}\right)_{T,V} \stackrel{LDA}{=}  \, - \frac{1}{2 V_T} \frac{1}{ \rho } \frac{\partial \langle n( {\bm \rho} ) \rangle}{\partial \rho} \, .
\end{equation}
can be computed in the way shown in Appendix \ref{appendix_derivative}. The LDA is a very good approximation for the density profile and only breaks down in the vicinity of the critical point~\cite{Wessel,Pollet_Criticality}. After averaging over the angular variable $\phi$ for radially symmetric lattices, ie. $ \{ L(\rho) , R(\rho) \} = \frac{1}{2\pi} \int d \phi \, \{ L({\bm \rho}), R({\bm \rho}) \}$, the temperature $T$ can be estimated from a least-square fit taking the measurement errors into account:
%
\begin{eqnarray}
\left( \sum_i \frac{R_i^2}{\Delta_{Li}^2} \right) - \left( \sum_i \frac{ L_i R_i }{ \Delta_{Li}^2 } \right) \hat{T} = \nonumber \\
- \left( \sum_i \frac{ L_i R_i }{ \Delta_{Ri}^2 } \right) \hat{T}^3 + \left( \sum_i \frac{ L_i^2 }{ \Delta_{Ri}^2 } \right) \hat{T}^4 
\end{eqnarray}
where $\Delta_{L}$ and $\Delta_{R}$ are the errors in $L({\bm \rho})$ and $R({\bm \rho})$,  respectively. 
Here, $\rho$ is discretized into bins with width corresponding to the experimental resolution. State of the art experiments allow us to set the binwidth to unity, even though we find that the scheme can tolerate a resolution of up to 5 sites (see Appendix \ref{appendix_resolution})~\cite{Chin2009, Greiner2009, Wuertz2009, Bakr2010, Sherson2010}. 
In the limit $\xi \rightarrow \infty$, this scheme reduces to the one of Zhou and Ho~\cite{Zhou09} in which the authors obtained an estimate for the temperature in an optical lattice of 1200 non-interacting fermions with 3\% error over 50 independent samples for a temperature $T/t=0.1$. 

For fermions,  both Ref.~\cite{Sanner} and Ref.~\cite{Muller} suggested that the FD theorem is an absolute thermometer in an harmonic trap without lattice, through careful normalization~\cite{Sanner} of $R_\infty ({\bm \rho})$,
However, for the small number $\mathcal{O}(10^5)-\mathcal{O}(10^6)$ of fermions in their experiments, the estimated temperature deviated from the time-of-flight (TOF) measurements by about 30\%~\cite{Muller}.

For bosonic optical lattices, our scheme is illustrated in figure \ref{fig:overall_FD} and \ref{fig:03} for the temperature of $T/t=1$, whereby a window size of $\xi=3$ is sufficent to capture (almost) all the correlations. Using $\xi=3$ as the standard for $T/t=1$ and higher temperatures, we show the number of independent measurements needed to estimate the temperature within 5\% error in Table \ref{table:nr_shot1}.  
\begin{table} [!htp]
\centering
\begin{tabular}{| c | c | c |}
\hline
System & \multicolumn{2}{|c|}{nr of shots} \\
\cline{2-3}
       & $\xi$=3  &  $\xi$=$\infty$ \\
\hline
$U/t=10$ , $T/t=1$  &  20  &  $\mathcal{O}(10^4)$ \\
$U/t=10$ , $T/t=3$  &  14  &  $\mathcal{O}(10^4)$ \\
$U/t=50$ , $T/t=1$  &  21  &  $\mathcal{O}(10^4)$ \\
$U/t=50$ , $T/t=3$  &  12  &  $\mathcal{O}(10^4)$ \\
\hline
\end{tabular}
\caption{Number of uncorrelated shots needed to obtain 5\% error in 3D ${}^{87}$Rb optical lattice experiments trapping 125,000 bosons with bin-width = 1.0 at $U/t = 10$ and $U/t = 50$. The variance reduction through window-sizing leads to orders of magnitudes improvement.
}
\label{table:nr_shot1}
\end{table}
The enormous variance reduction through window-sizing turns the FD thermometry scheme into a feasible tool for ultracold bosonic optical lattice experiments.
With 20 independent shots uniformly distributed over 20\% spread in $T$ and 1\% in $N$, this scheme remains applicable (see Appendix \ref{appendix_fluctuations}). 
\begin{figure}
\centering
\includegraphics[width=7.5cm]{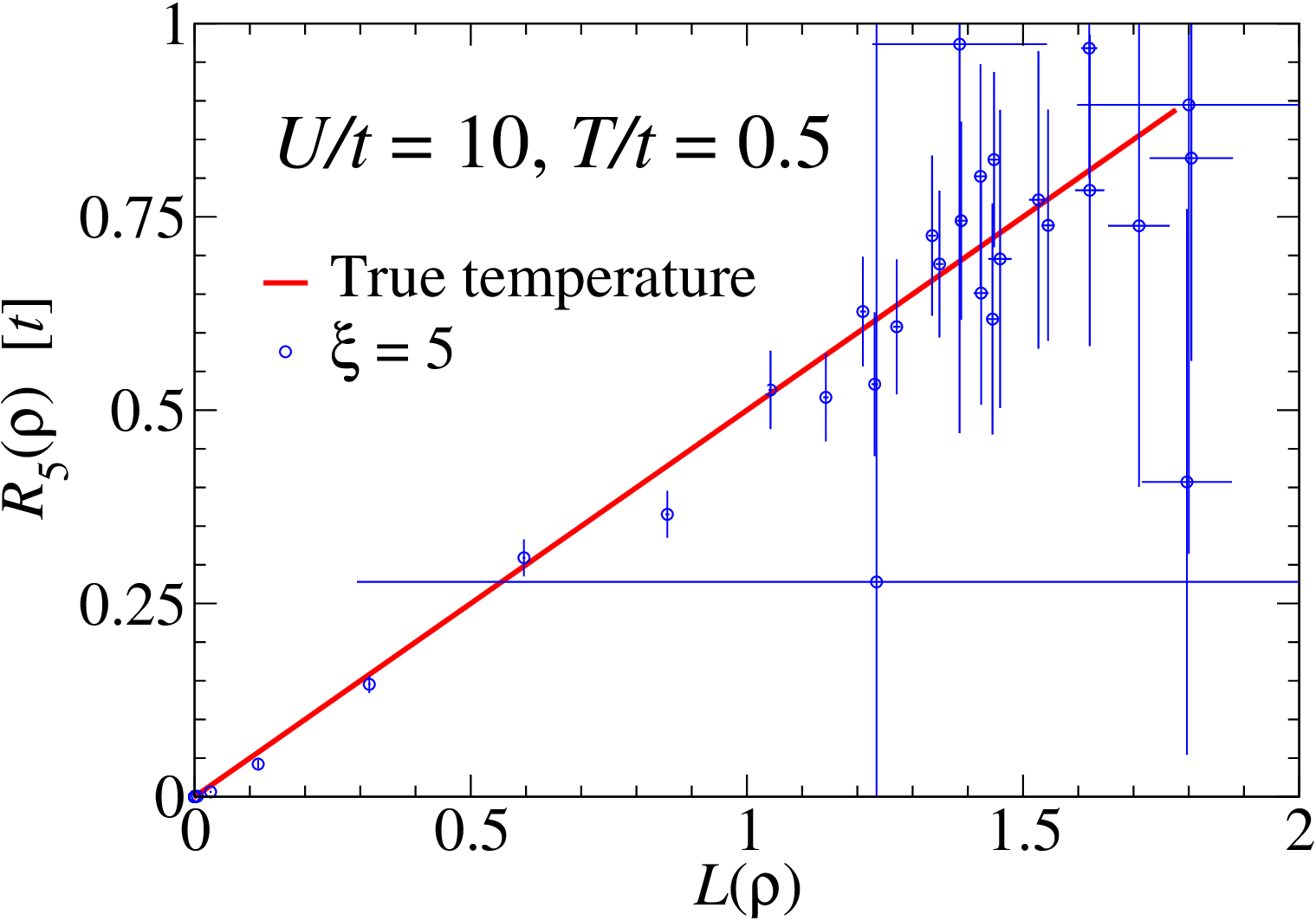}
\caption
{
Fluctuation-dissipation thermometry scheme at slightly lower temperature. We take a 3D bosonic ${}^{87}$Rb optical lattice system with parameters $N = 125,000$, $U/t = 10$, $T/t = 0.5$, $\xi = 5$,  and we average over 100 independent measurements obtained from a QMC simulation.
}
\label{fig:LR_0.5}
\end{figure}
At lower temperature, the scheme remains valid although a larger $\xi$ is needed, and the statistical noise will inevitably grow. An example of T/t = 0.5 is shown in figure \ref{fig:LR_0.5} where the correlations are effectively captured by a window size of $\xi=5$, and 100 independent measurements are required to attain an accurate temperature estimate with 5\% error.

\begin{figure}
\centering
\includegraphics[width=4cm]{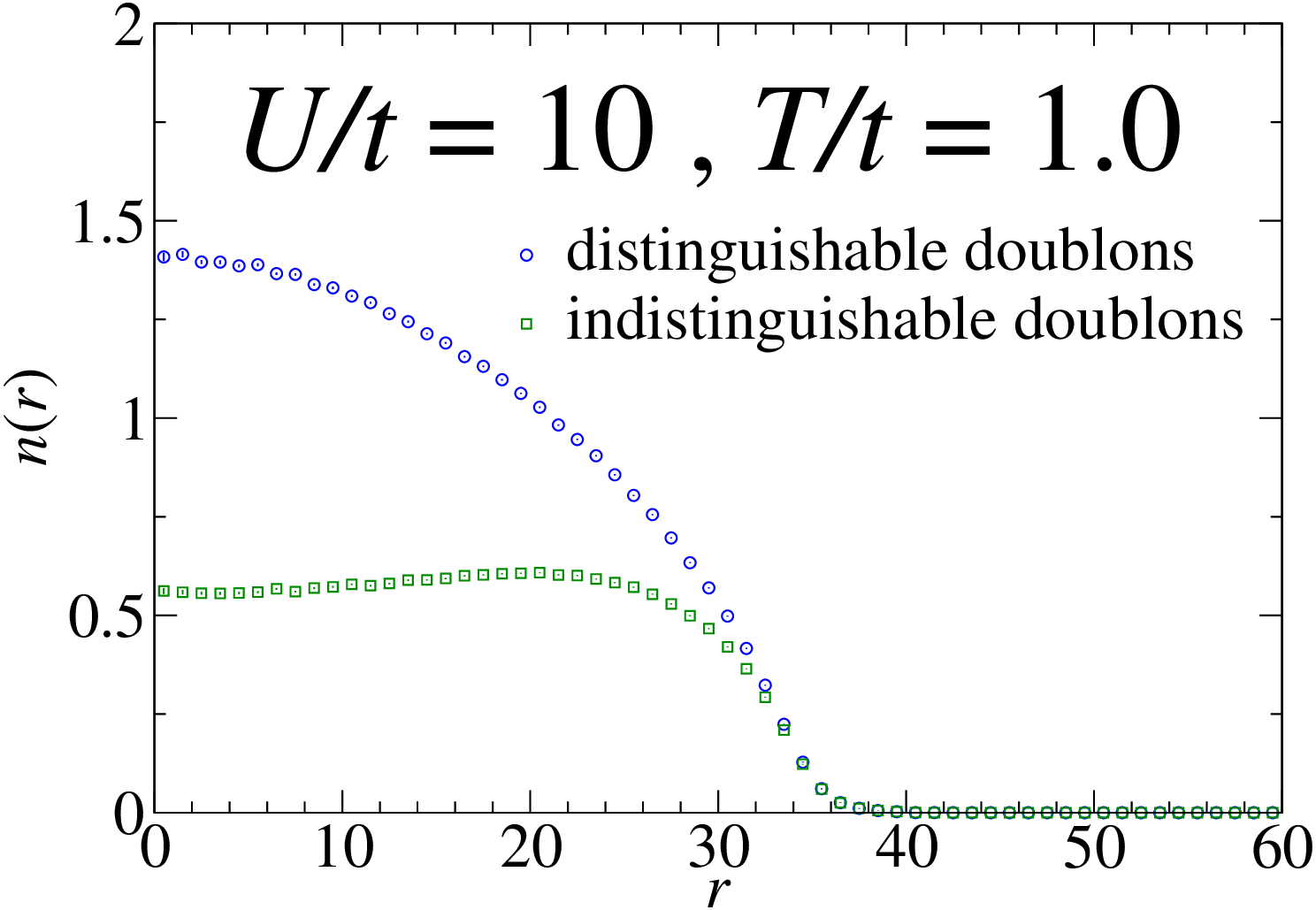}
\hfill
\includegraphics[width=4cm]{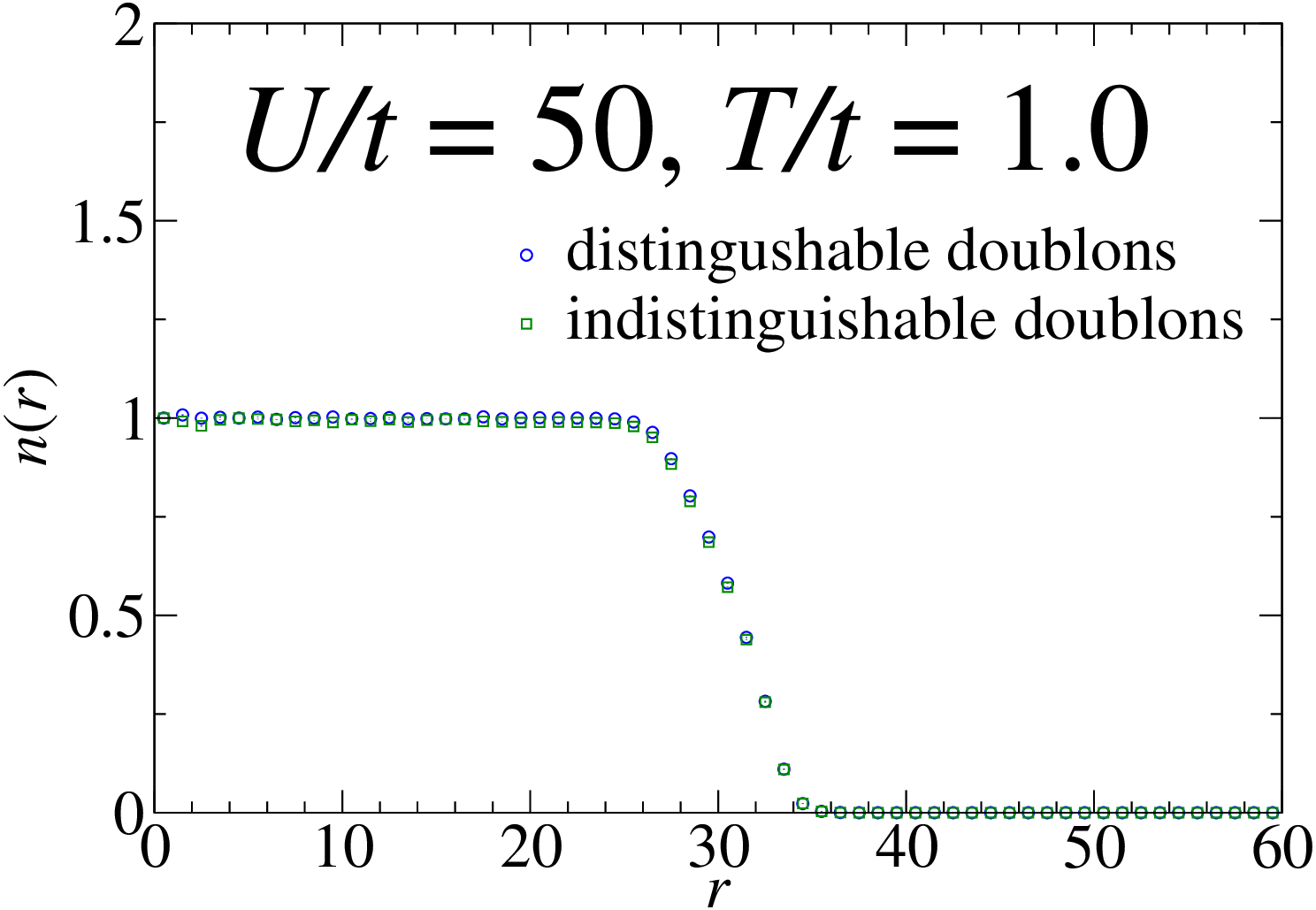}
\hfill
\includegraphics[width=4cm]{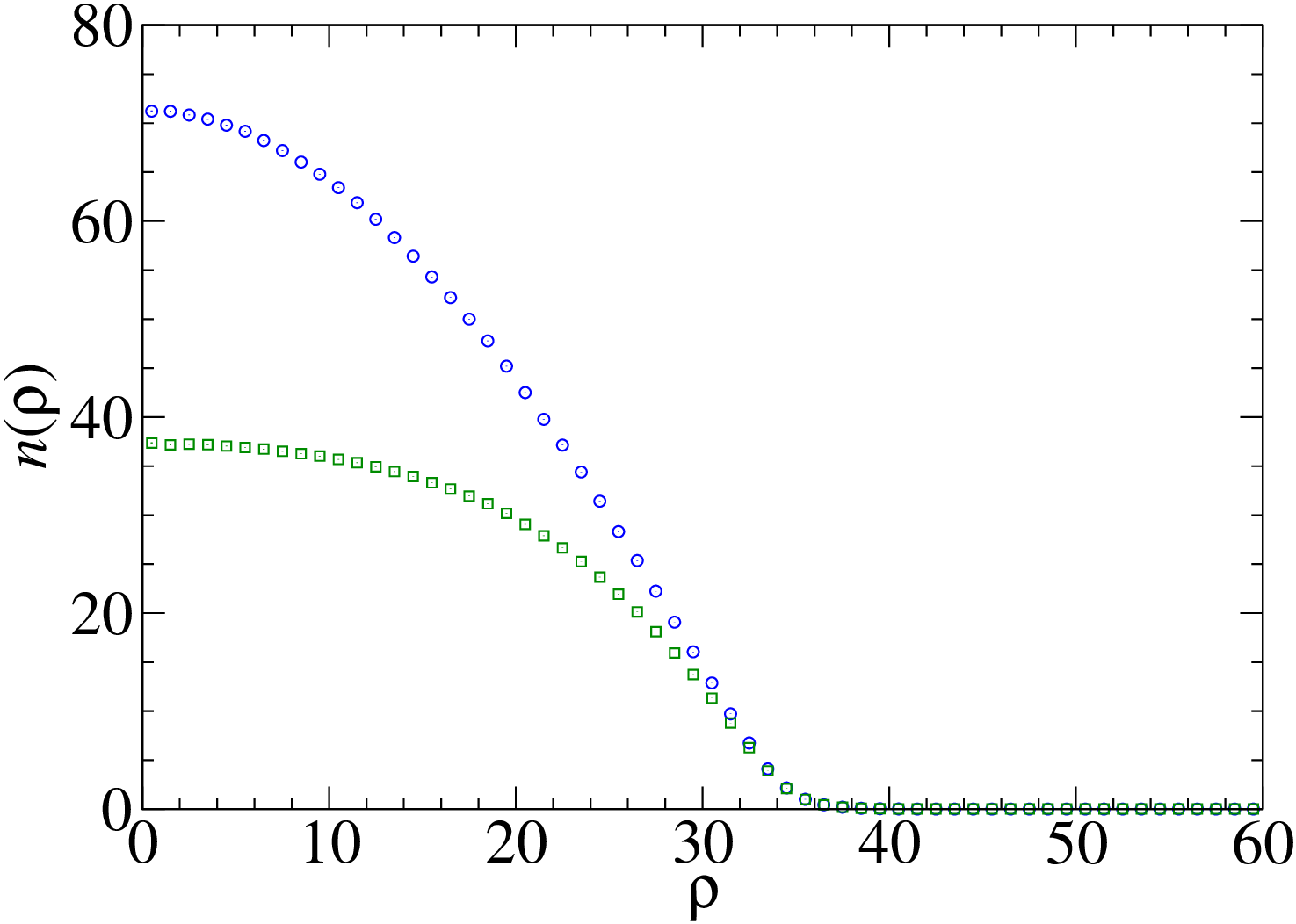}
\hfill
\includegraphics[width=4cm]{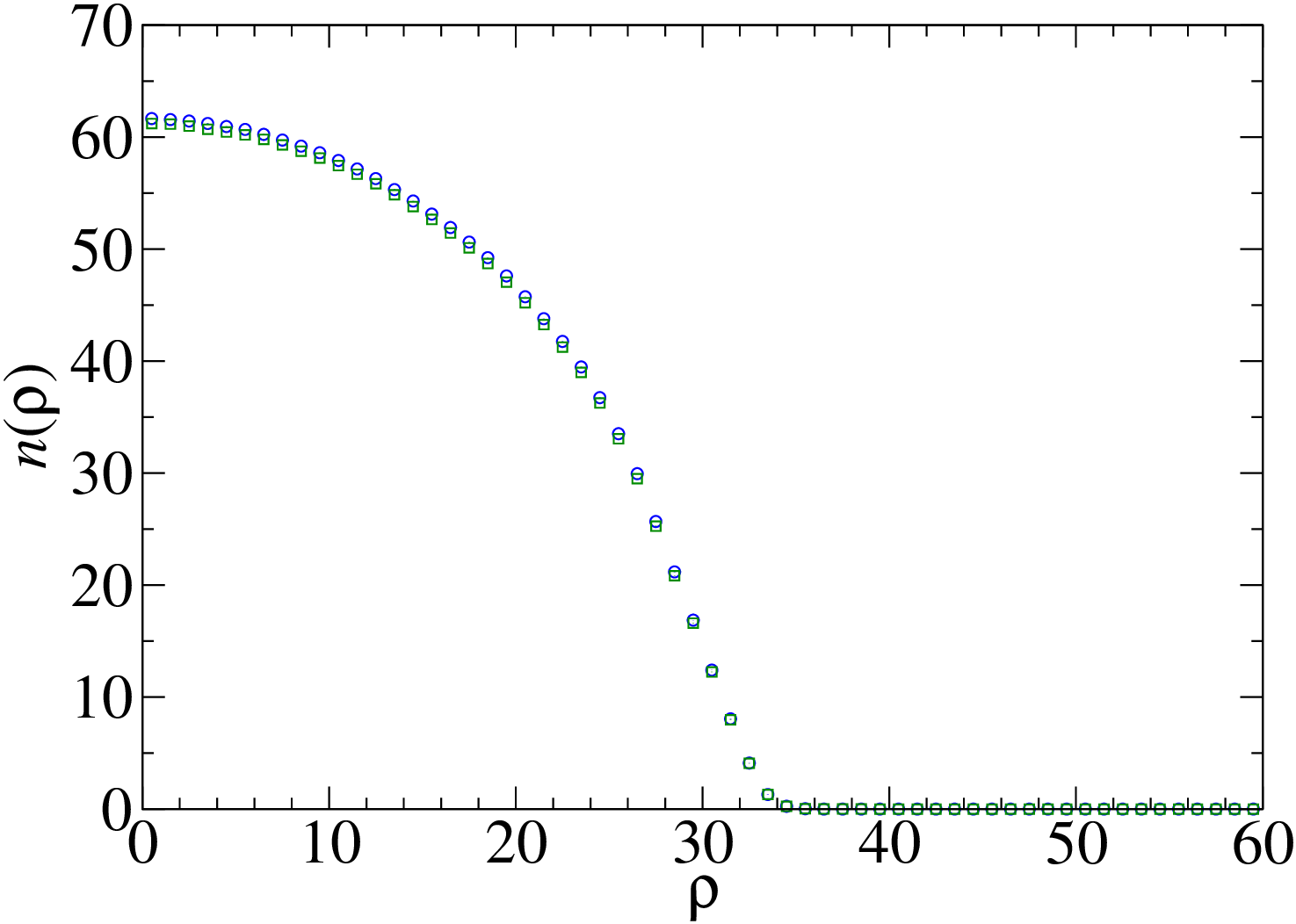}
\hfill
\includegraphics[width=4cm]{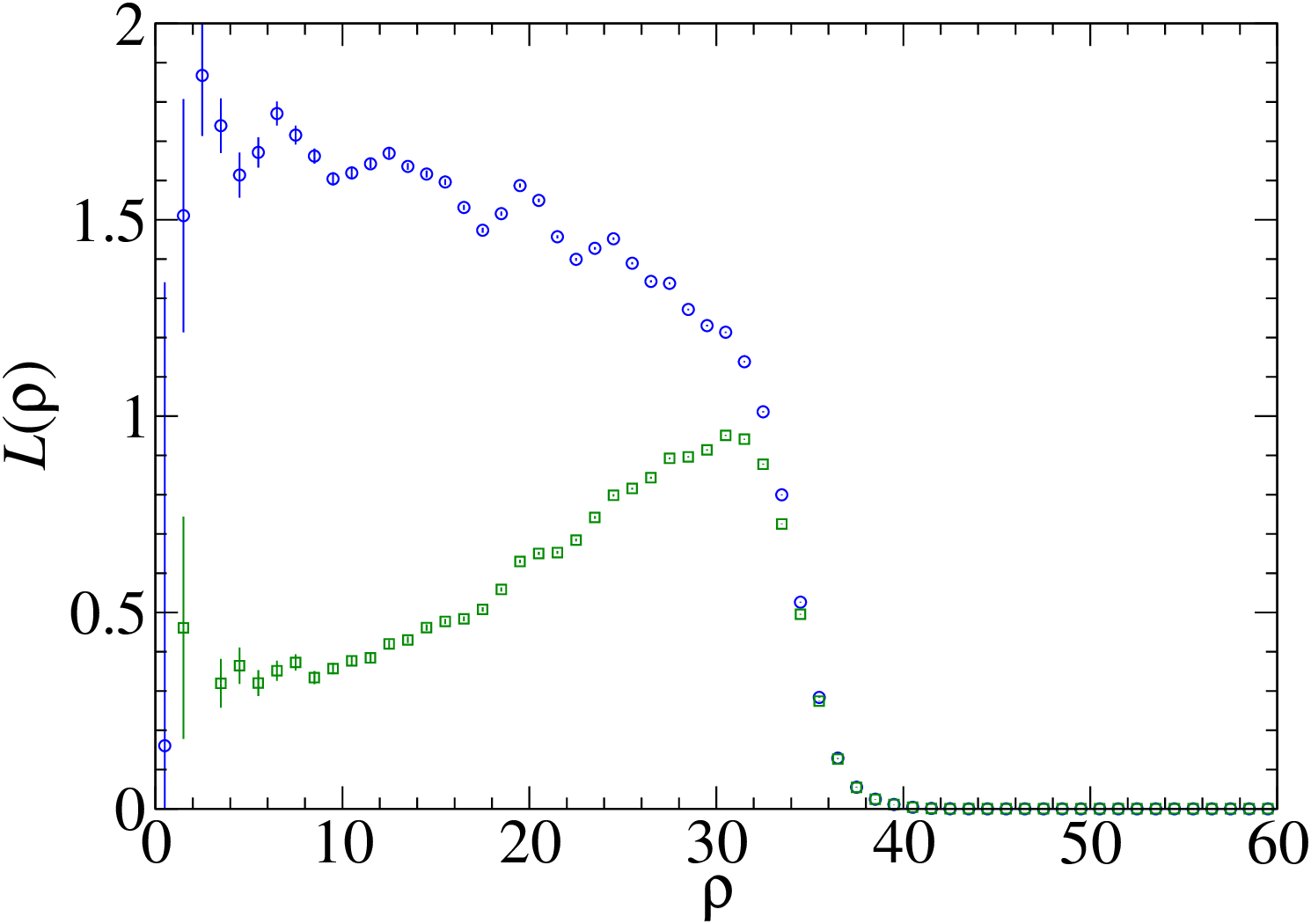}
\hfill
\includegraphics[width=4cm]{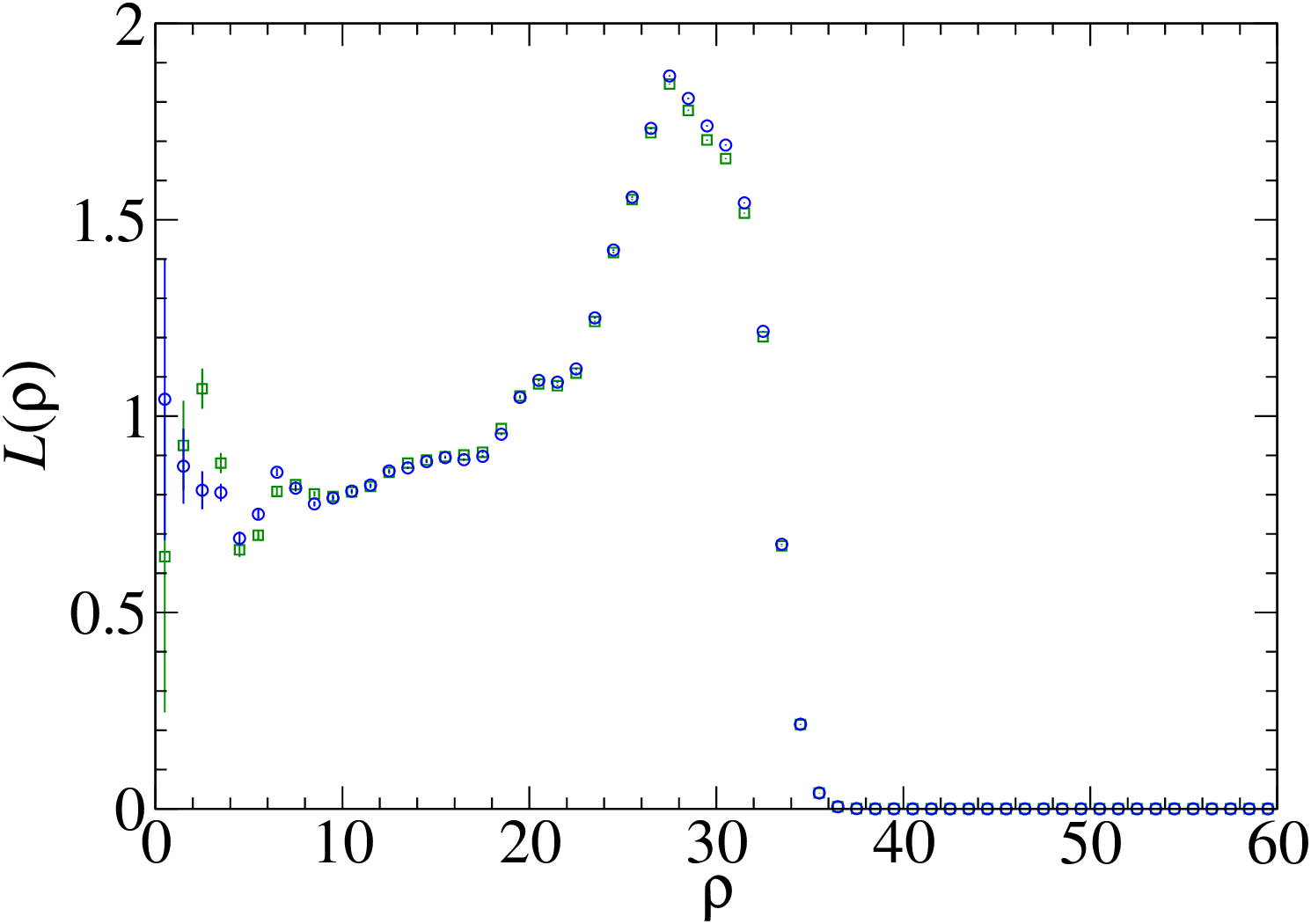}
\hfill
\includegraphics[width=4cm]{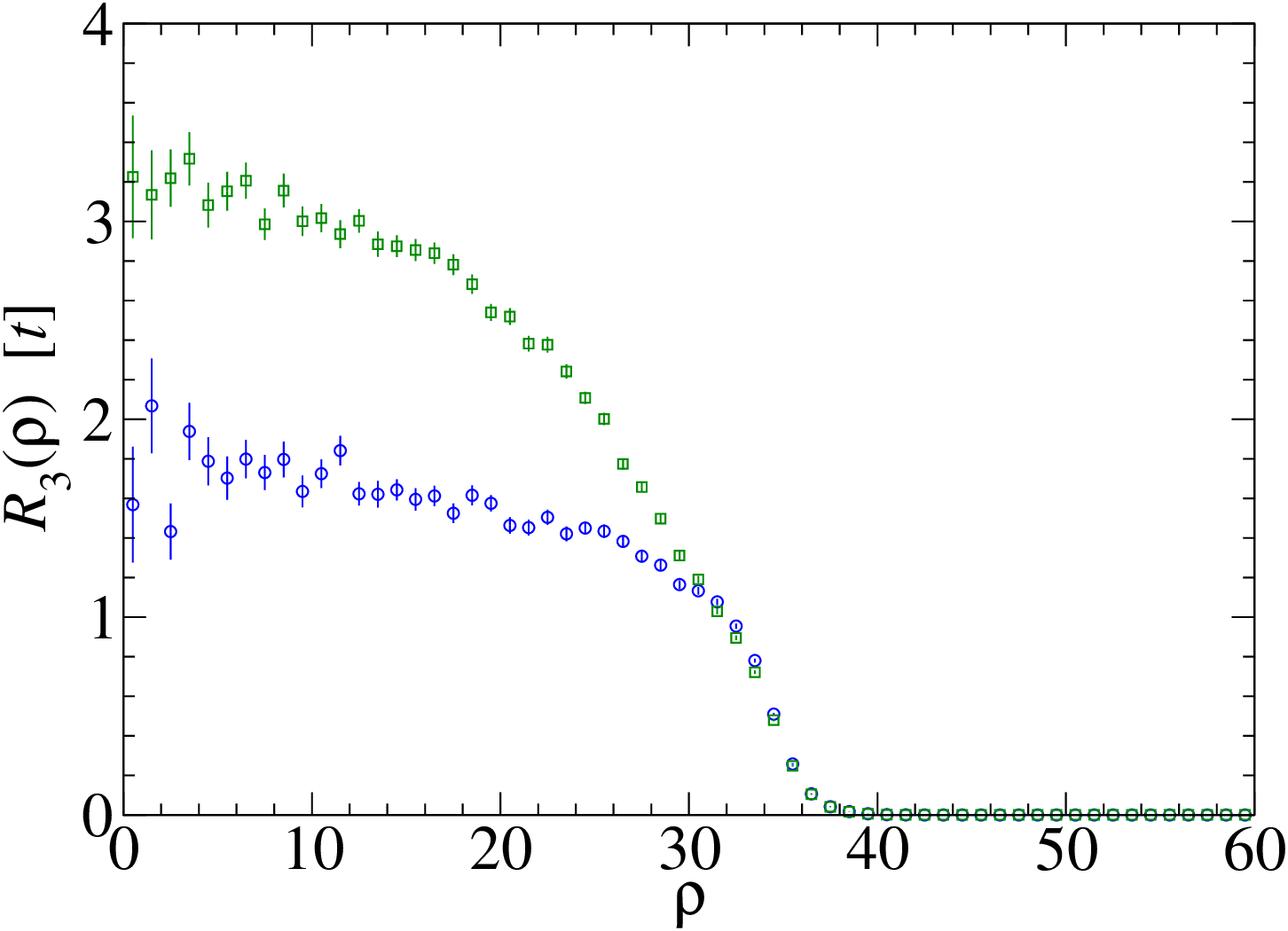}
\hfill
\includegraphics[width=4cm]{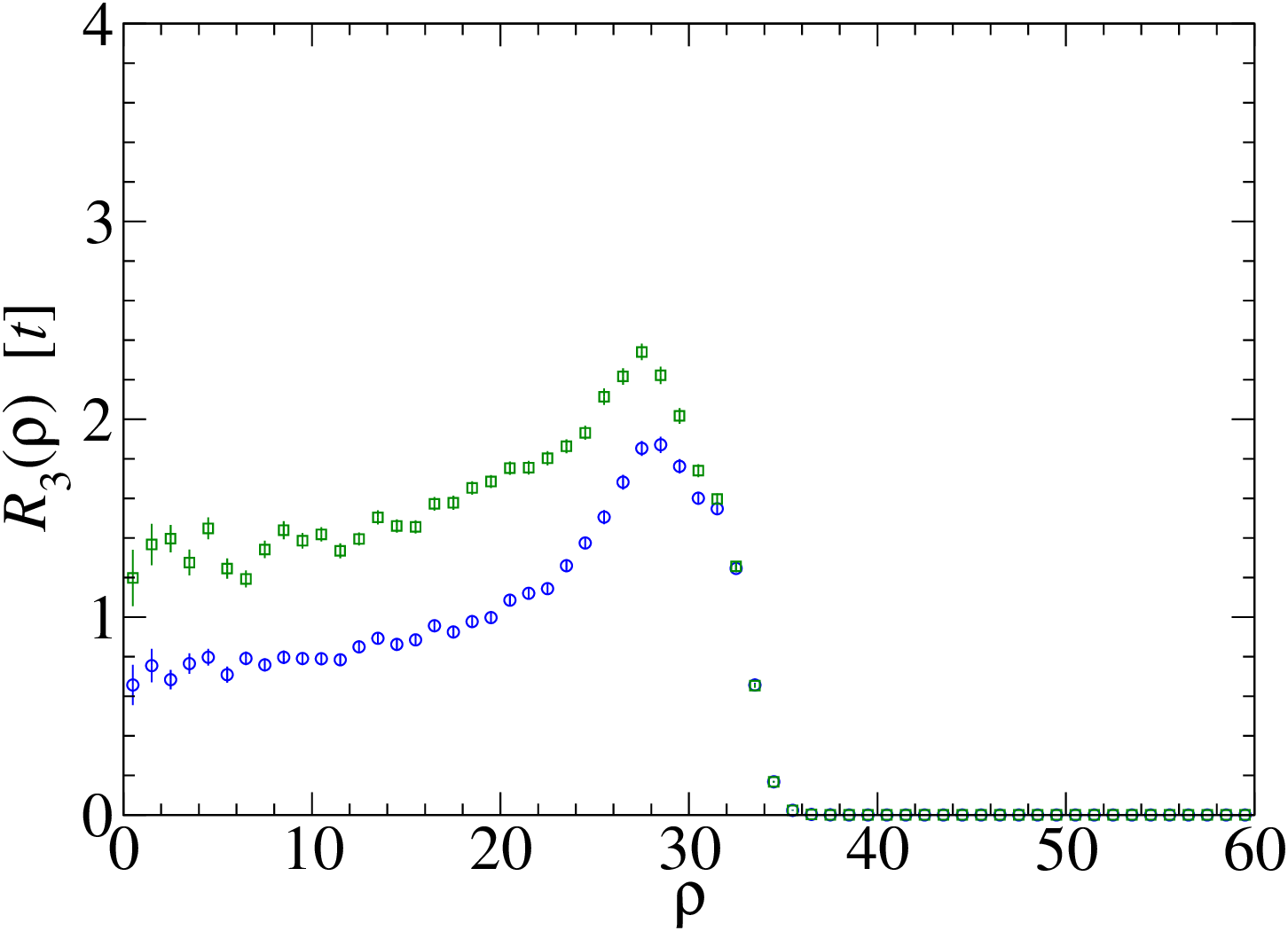}
\hfill
\includegraphics[width=4cm]{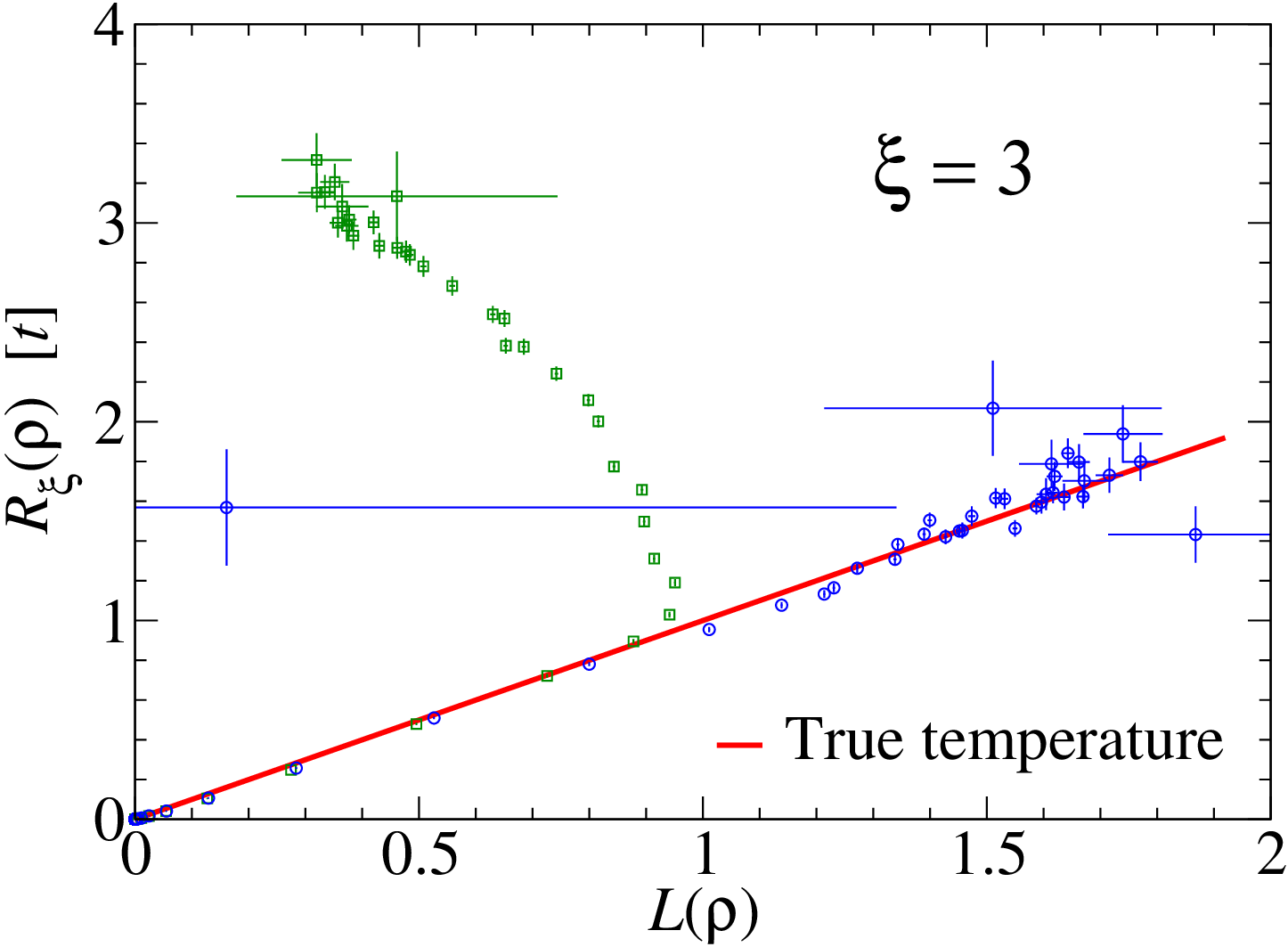}
\hfill
\includegraphics[width=4cm]{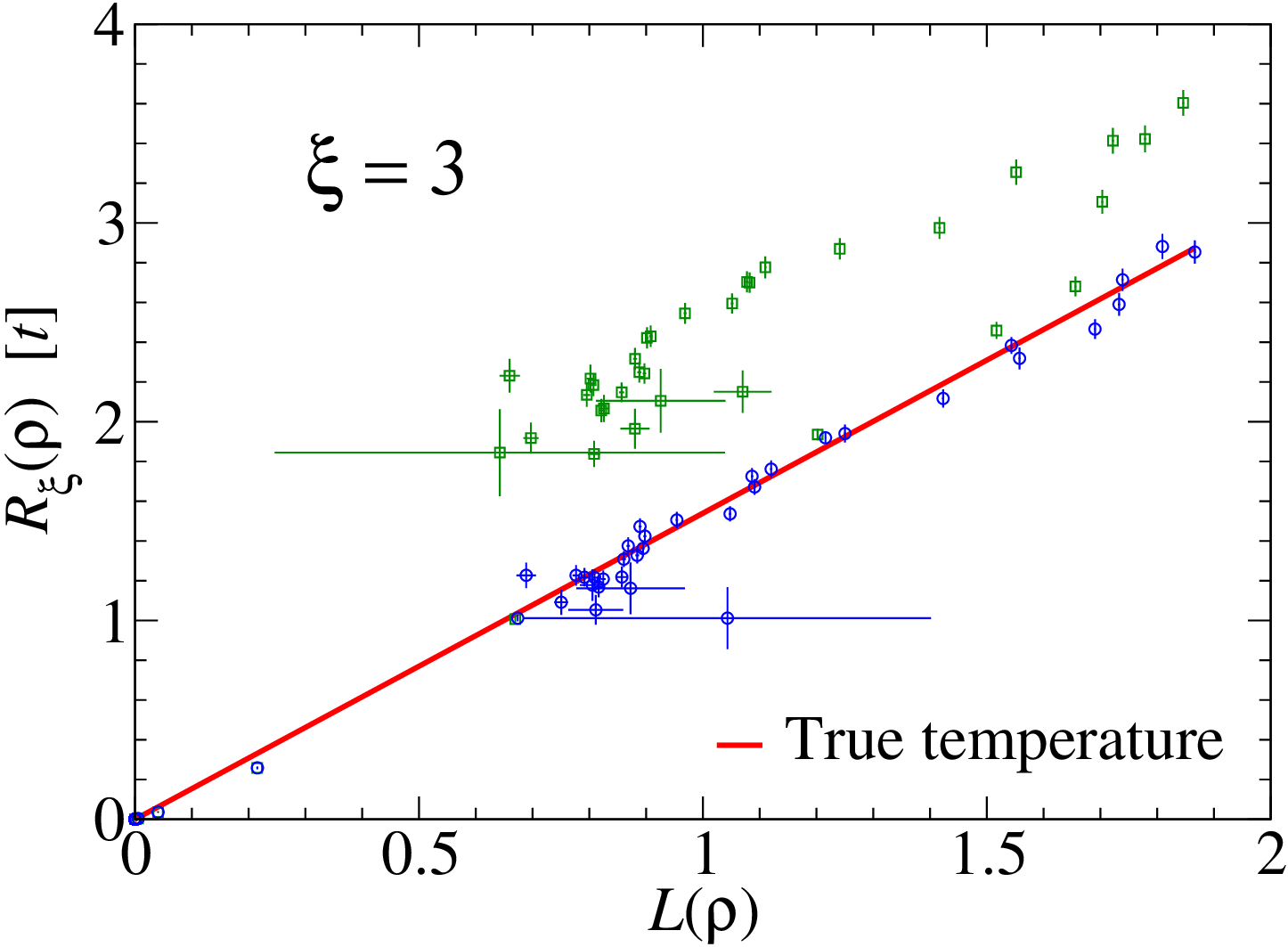}
\caption
{
Fluctuation-dissipation thermometry scheme in the presence of doublon-hole indistinguishability. From top to bottom are shown : 1) cross-section density n(r) , 2) column integrated density n($\rho$) , 3) dissipation term L($\rho$) , 4) fluctuation term ($\xi=3$) $R_3(\rho)$ , 5) $L(\rho)$-$R_3(\rho)$ relationship.
We take a 3D bosonic ${}^{87}$Rb optical lattice system with $N = 125,000$, and we average over 1000 independent measurements obtained from a QMC simulation.
In the left column the parameters are  $U/t = 10$ , $T/t = 1$; and in the right column $U/t = 50$ , $T/t = 1$.
Blue circles (green squares) show the curve where doublons can (cannot) be distinguished from holes.
}
\label{fig:even_odd}
\end{figure}
Current optical lattice experiments using fluorescence techniques can only measure the parity (even/odd) of the occupation number per site.
This affects the FD thermometry scheme, as illustrated in figure \ref{fig:even_odd}, but through selection of those points in the $L(\rho)$-$R_\xi(\rho)$ that are on the linear slope, we could still obtain an acceptable estimate of the temperature. Deep in the edges, the number of doublons is very low compared to the number of holes due to the low overall density and the high potential energy cost of creating a doublon.

%
\begin{table} [!htp]
\centering
\begin{tabular} {| c | c |}
\hline
System              & estimated temperature [t]  \\
\hline
$U/t=10$ , $T/t=1 $ &  0.985 $\pm$ 0.008  \\
$U/t=50$ , $T/t=1 $ &  1.003 $\pm$ 0.012  \\
\hline
\end{tabular}
\caption{The estimated temperature  for a 3D bosonic ${}^{87}$Rb optical lattice system in the presence of doublon-hole indistinguishability. The parameters are $N = 125,000$, bin-with = 1.0, $\xi$ = 3 at $U/t = 10$,$50$ and $T/t = 1$. We average over 1000 uncorrelated measurements obtained from a QMC simulation.}

\label{table:fitted_temp_even_odd}
\end{table}

%% file: second_order_high_temperature_thermometry.tex
\section{Second-order high temperature expansion}

In this section we propose an alternative scheme for thermometry based on second-order high-temperature series expansion (HTE2). 
Deep enough in the edges, there will always be a normal region for any temperature and interaction strength where the system is well described by second order high-temperature series expansions.
One advantage of this scheme is that it works not only for integrated column densities but also for density profiles measured only in 2D cross sections of a 3D system \footnote{Here the FD based scheme fails if the cross section is thinner than the correlation length}. In addition this scheme allows determination of the chemical potential.

In practice, a single shot will suffice to extract the quantities of interest. When HTE2 applies, it also gives a foundation for the limit on the small window size $\xi$ for the former fluctuation dissipation thermometry scheme (indeed, there is no point in applying the FD scheme whenever HTE2 works). It is possible to use higher-order schemes than HTE2, but we find that the gain is minimal compared to the additional effort.
However, it may be that
the density in the edges is so low that it cannot be measured because of the low signal-to-noise ratio, which will happen for very low temperatures. In this section, we wish to make these ideas more quantitative.

\begin{figure}[!htp]
\includegraphics[width=7cm]{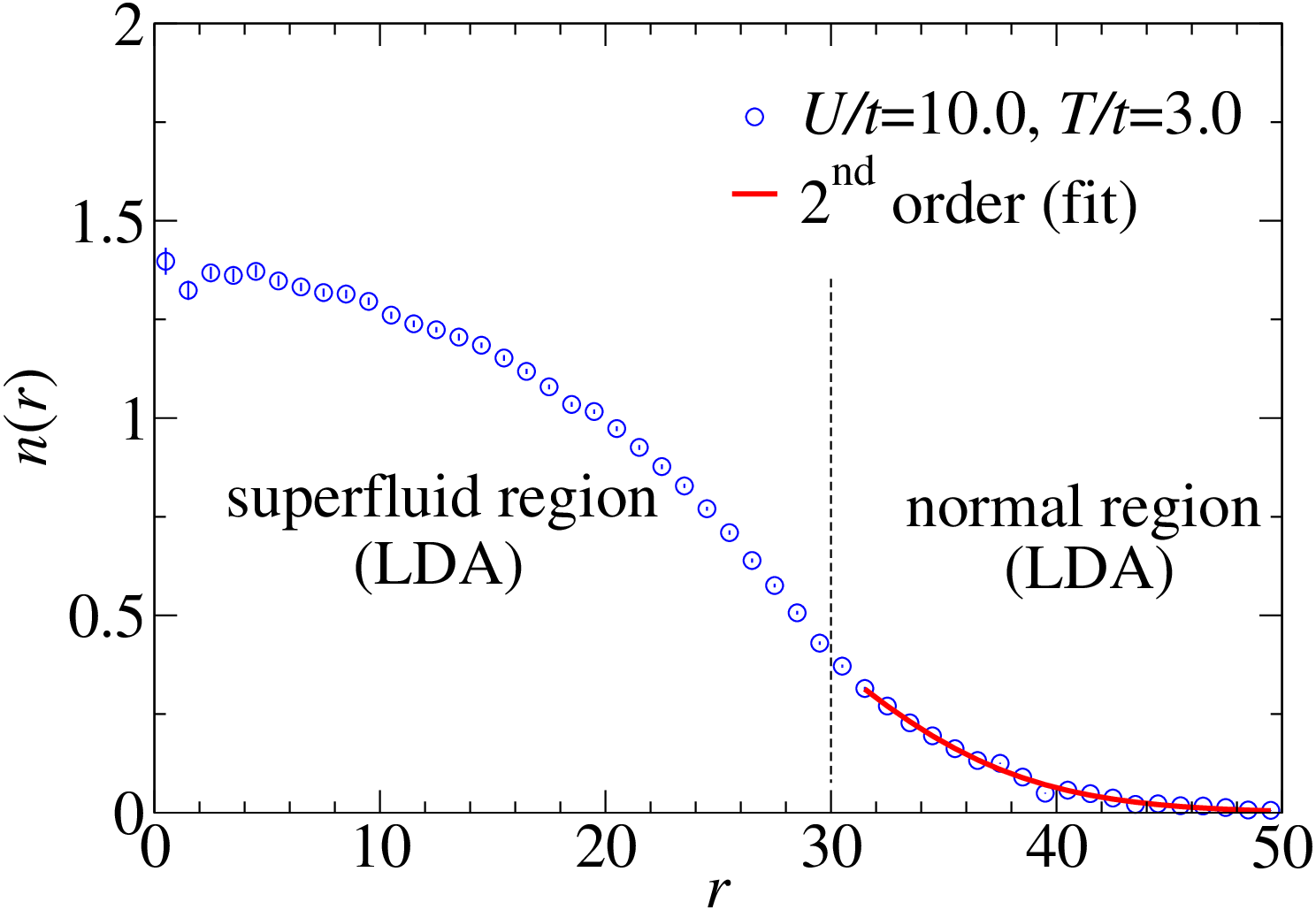}
\caption
{
Second order high temperature series expansion thermometry scheme for a bosonic optical lattice system. 100 cross-sectional density measurements are used to estimate the temperature and chemical potential. The system consists of a 3D optical lattice with ${}^{87}$Rb atoms with parameters  $N = 125,000$, $\mu/t = 4.835$, $U/t = 10$, $T/t = 3$.  The blue circles are data averaged over 100 measurements obtained from a QMC simulation; the red line is a  least-square fit over the normal region where $\mu_{\rm fit}/t = 5.244$ and $T_{\rm fit}/t = 2.820$.
}
\label{fig:mixed_SF_normal}
\end{figure}

In the absence of correlations the density $\langle n_i^{(0)} \rangle$ is captured by the zeroth-order high temperature expansion theory and given by
\begin{equation}
\label{density0}
\langle n_i^{(0)} \rangle = \frac{1}{Z_i^{(0)}} \sum_{\{ n_i \}} n_i e^{-\beta (D_i - \mu_i n_i )}  \,\, ,
\end{equation}
where the zeroth order partition function $Z_i^{(0)}$ is 
\begin{equation}
\label{partition0}
Z_i^{(0)} = \sum_{\{ n_i \}} e^{-\beta ( D_i - \mu_i n_i) } \,\, ,
\end{equation}
and the onsite diagonal energy $D_i$ is
\begin{equation}
\label{diagonal_subHamiltonian}
D_i = \frac{U}{2} n_i(n_i-1)  \,\, .
\end{equation}
\begin{figure}[!htp]
\includegraphics[width=7cm]{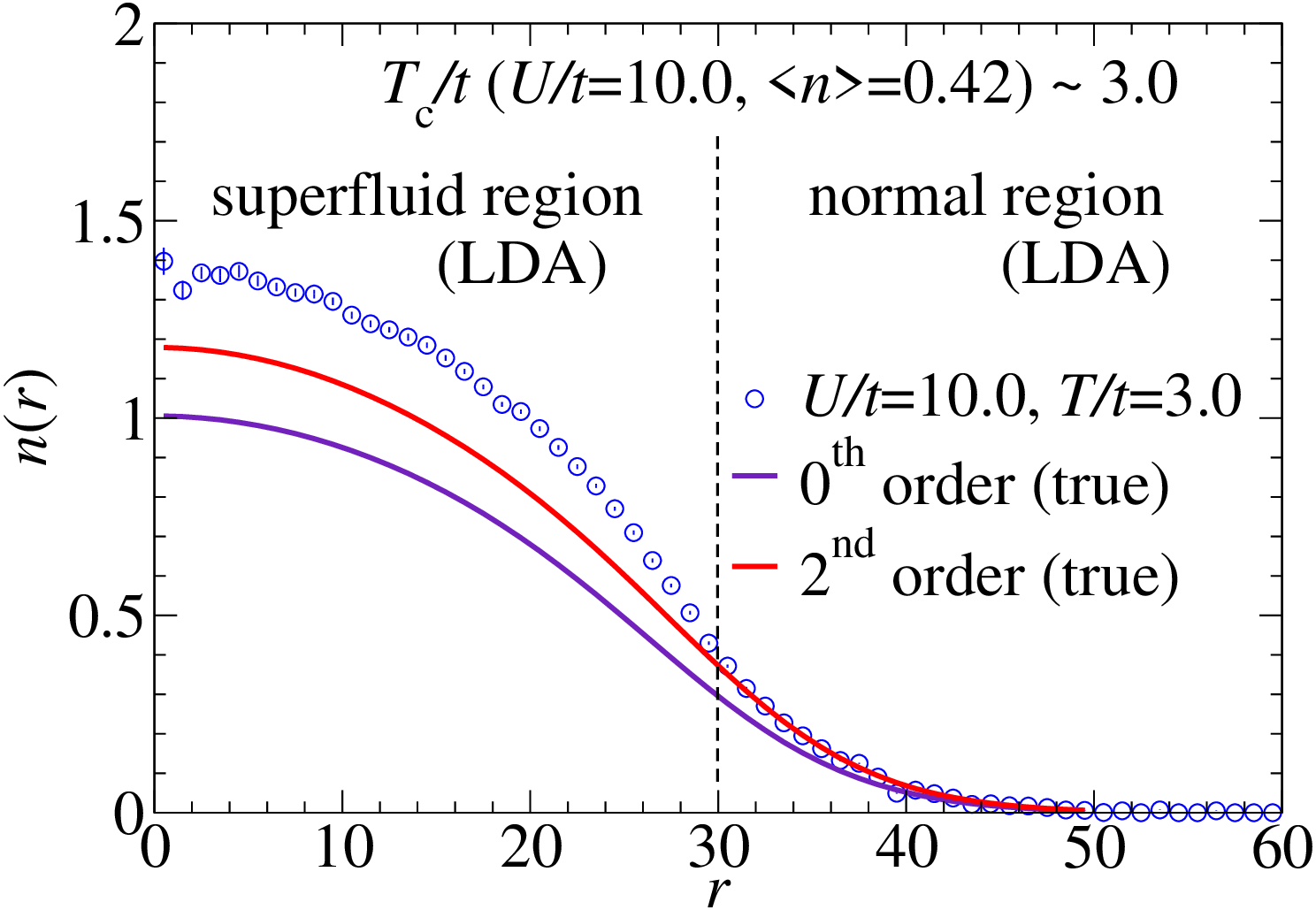}
\caption
{
Illustrating the concept of wing thermometry for a 3D bosonic ${}^{87}$Rb optical lattice system, {\it ie.,} describing the normal state by high temperature series expansions. Blue circles: In-situ density profile obtained from 100 uncorrelated measurements obtained by a QMC simulation with parameters
$U/t = 10$ , $T/t = 3$ , $N = 125,000$. The superfluid-normal phase boundary occurs at the density $\langle n \rangle=0.42$ or chemical potential $\mu / t = -2.75$. 
The $2^{nd}$ order series captures all the physics in the normal regime, whereas the $0^{th}$ order has a very small validity range.
} 
\label{fig:densityorder0}
\end{figure}
The zeroth order expansion, suggested in Ref.~\cite{Zhou_Kato},  does a rather poor job in describing the edges of the system, as illustrated in Fig.~\ref{fig:densityorder0}. 
However, sufficient accuracy over a wide density range in present-day experiments is found by emplying the second order (up to $(\beta t)^2$)  partition function, 
\begin{equation}
Z = Z^{(0)} ( 1 + Z^{(2)} ) \, ,
\end{equation}
with
\begin{eqnarray}
Z^{(2)} &=& \sum_{\langle i,j \rangle} \frac{(-\beta t)^2}{Z_i^{(0)} Z_j^{(0)}}  \left[ \sum_{\{ n_i,n_j \} }^{(-,+)} \, n_i n_j^\delta e^{-\beta ( D_i + D_j - \mu_i n_i - \mu_j n_j)} \Gamma_{ij}^\delta\right. \nonumber \\
        & & \left. + \sum_{ \{n_i,n_j \} }^{(+,-)} \, n_i^\delta n_j e^{-\beta ( D_i + D_j - \mu_i n_i - \mu_j n_j)} \Gamma_{ji}^\delta \right] \, ,
\end{eqnarray}
where $\langle i, j \rangle$ denotes  the sum over nearest neighbors, $(-,+)$ stands for $(n_i^\delta = n_i - 1,n_j^\delta = n_j + 1)$ and vice versa for $(+, -)$.
$\Gamma_{ij}^\delta$ is defined as
\begin{equation}
\Gamma_{ij}^\delta = \frac{1 - e^{\beta \gamma_{ij}^{\delta}}}{\left(\beta \gamma_{ij}^{\delta}\right)\left(\beta\gamma_{ji}\right)} - \frac{1 - e^{\beta \left(\gamma_{ij}^{\delta}+\gamma_{ji}\right) }}{\left(\beta\gamma_{ij}^{\delta}+\beta\gamma_{ji}\right)\left(\beta\gamma_{ji}\right)},
\end{equation}
and we also introduce $\chi_{ij}^\delta$,
\begin{eqnarray}
\chi_{ij}^\delta &=& \frac{e^{\beta \gamma_{ij}^{\delta}}}{\left(\beta \gamma_{ij}^{\delta}\right)\left(\beta\gamma_{ji}\right)} + \frac{1 - e^{\beta \left(\gamma_{ij}^{\delta}+\gamma_{ji}\right) }}{\left(\beta \gamma_{ij}^{\delta}+\beta\gamma_{ji}\right)\left(\beta\gamma_{ji}\right)^2} \nonumber \\
 & & - \frac{\left(1 - e^{\beta \gamma_{ij}^{\delta}}\right)\left(\beta \gamma_{ij}^{\delta} - \beta\gamma_{ji}\right)}{\left(\beta \gamma_{ij}^{\delta}\right)^2 \left(\beta\gamma_{ji}\right)^2} ,
\end{eqnarray}
with $\gamma_{ij}^{(\delta)} = U(n_i^{(\delta)} - n_j^{(\delta)} + 1) + (\mu_i - \mu_j)$.
The temperature is found by fitting the cross-sectional experiment in-situ density measurement against the  density found in second order high temperature series expansions (HTE2) , 
\begin{eqnarray}\label{HTE2_density}
&& \langle n_i \rangle = \langle n_i^{(0)} \rangle \,+\, \sum_{\langle i, j \rangle} \frac{(-\beta t)^2}{Z_i^{(0)} Z_j^{(0)}}\times \\
& &\left[ \sum_{ \{n_i,n_j \} }^{(-,+)} \left( \delta n_i + \frac{\chi_{ij}^\delta}{\Gamma_{ij}^\delta} \right)  n_i n_j^\delta e^{-\beta \left( D_i + D_j -\mu_i n_i - \mu_j n_j \right)} \Gamma_{ij}^\delta \right. \nonumber \\
& &\left. + \sum_{ \{n_i,n_j \} }^{(+,-)} \left( \delta n_i - \frac{\chi_{ji}^\delta}{\Gamma_{ji}^\delta} \right) n_i^\delta n_j e^{-\beta \left( D_i + D_j -\mu_i n_i - \mu_j n_j \right)} \Gamma_{ji}^\delta \right] \, \nonumber 
\end{eqnarray}
where $\delta n_i = n_i - \langle n_i^{(0)} \rangle$. 
HTE2 gives a fairly accurate description of the edges in present experiments. In practice, one should fit from some distance $r_1$ till the end of the trap, and vary $r_1$ in order to find the range of applicability of the HTE2 scheme, see Fig.~\ref{fig:mixed_SF_normal}.

\begin{figure}[!htp]
\includegraphics[width=7cm]{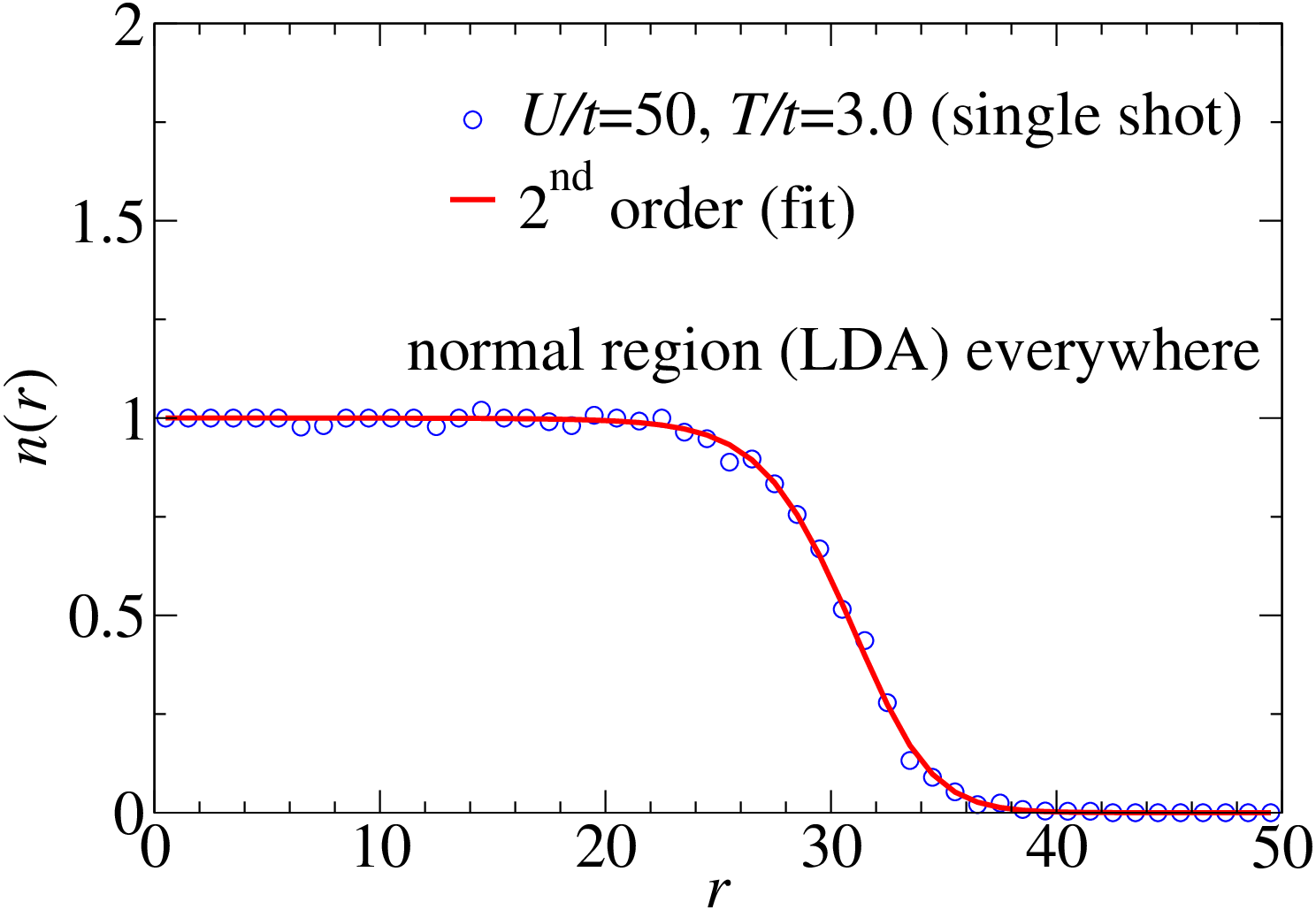}
\caption
{
Second order high temperature series expansion thermometry scheme for a bosonic system that is entirely in the normal phase. No more than a single shot of cross-sectional density is needed to estimate the temperature and chemical potential within 10\% accuracy. We take a 3D optical lattice system with bosonic ${}^{87}$Rb atoms and parameters $N = 125,000$, $\mu/t = 25.97$, $U/t = 50$, $T/t = 3$. The blue circles are obtained from a single measurement in a QMC simulation; and the red line is a least-square fit over the entire normal region where $\mu_{\rm fit}/t = 25.92$ and $T_{\rm fit}/t = 2.824$ nK.
}
\label{fig:singleshot}
\end{figure}
If the entire regime of the bosonic optical lattice is in the normal phase, one would require no more than a single experimental density measurement to extract a reliable estimate of the temperature and chemical potential. This is shown in figure \ref{fig:singleshot}.

%% file: discussion_acknowledgement.tex
\section{Conclusions and Acknowledgements}

We have discussed how single-site resolution detection tools can be used to obtain the equation of state and/or the temperature in trapped ultracold gases in an optical lattice.
We analyzed whether the fluctuation-dissipation theorem can be used to extract temperature for present experiments when the LDA is valid~\cite{Zhou09} over the entire tap. 
Taking advantage of the fact that the density-density correlation length is short away from the critical region (also in the superfluid phase), a few dozen measurements are sufficient in order to extract the temperature reliably and accurately under the condition that the shape of the chemical potential landscape is known, the system is in thermodynamic equilibrium, and that the local density approximation holds. 

Using more theoretical input, such as density profiles obtained in a second order high temperature expansion the temperature can be obtained from the normal edges sometimes already with a {\em single} measurement. In cases where the normal region on the edge is too  narrow one can either go to higher order in the high temperature expansion or experimentally shape the trap to obtain a wider normal region.

The ALPS scheduler and alea libraries \cite{ALPS} were used for parallelization and Monte Carlo data analysis.
The simulations were performed on the Brutus cluster of ETH Z\"urich, and the entire workflow was carried out in the Vistrails framework.
We thank M. Cheneau, T-L. Ho, N. V. Prokof'ev, B. V. Svistunov,  L. Tarruel, D-W. Wang, and Q. Zhou for useful discussions.
We acknowledge financial support from the Swiss National Science Foundation and hospitality of the Aspen Center for Physics.

%% file: appendix.tex
\appendix

\section{Numerical approximation to the density derivative}
\label{appendix_derivative}

In the dissipation term
\begin{equation}
L(\rho) = - \frac{1}{2 V_T} \frac{1}{ \rho } \frac{\partial \langle n( \rho ) \rangle}{\partial \rho}  \, ,
\end{equation}
the derivative needs to be taken numerically, which always involves an approximation.
The most direct way is to use central differences, but to improve on the quality of data one could do the following: first perform a local parabolic fitting on each $\langle n(\rho) \rangle$ density bin over a fitting range of (2k+1) density bins, {\it ie.}, over the interval  $\left[ \langle n(\rho-k) \rangle , \langle n(\rho+k) \rangle \right]$ before taking the spatial derivative analytically from the fitted density profile. In practice, we find that the fluctuation dissipation thermomtery scheme is unaffected by choosing different numerical approximations to calculating the derivative. However, large density gradients always result in systematic errors. 

\section{In-situ density imaging with few-sites resolution}
\label{appendix_resolution}

\begin{figure*}
\centering
\includegraphics[width=8.5cm]{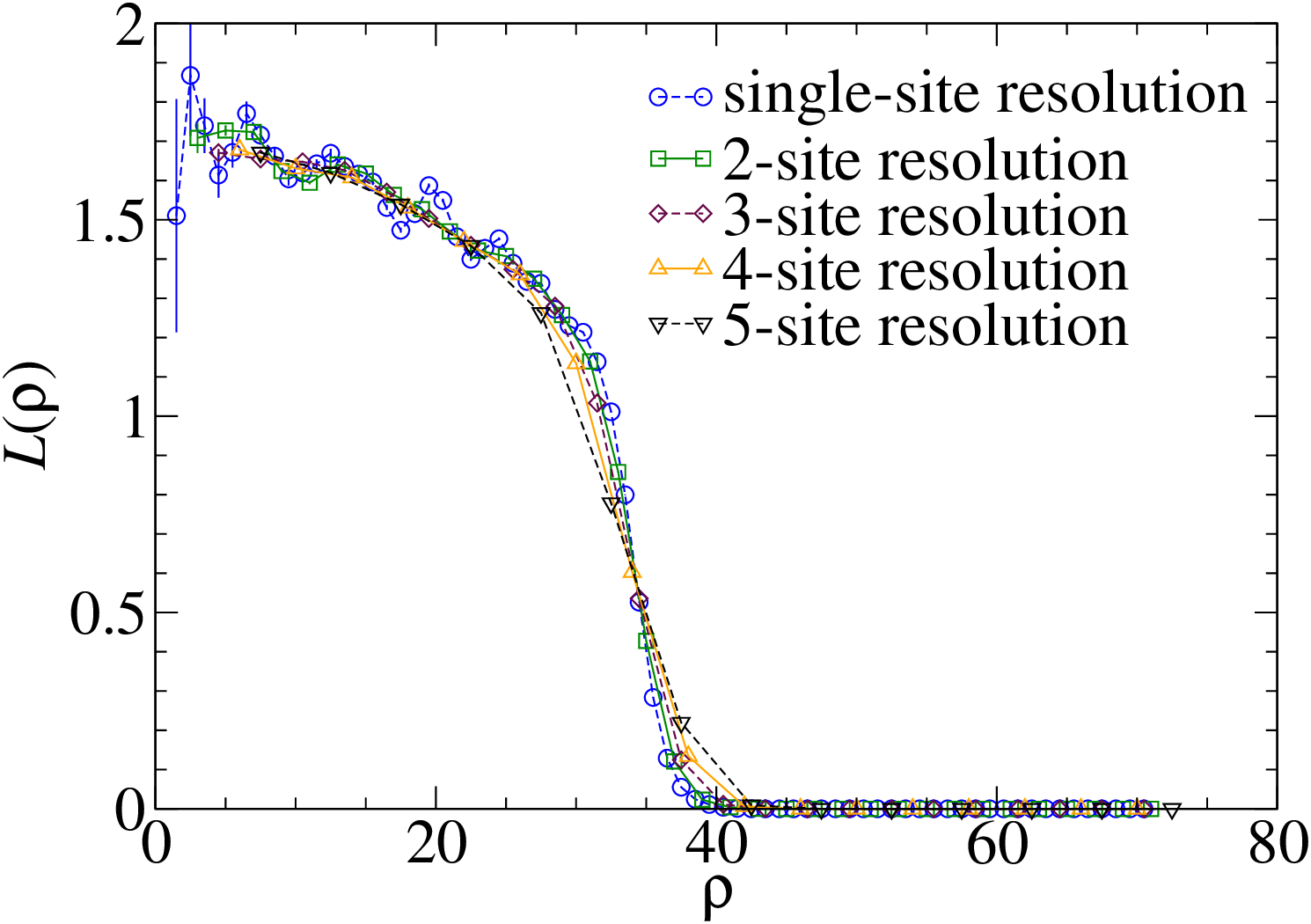}
\hfill
\includegraphics[width=8.5cm]{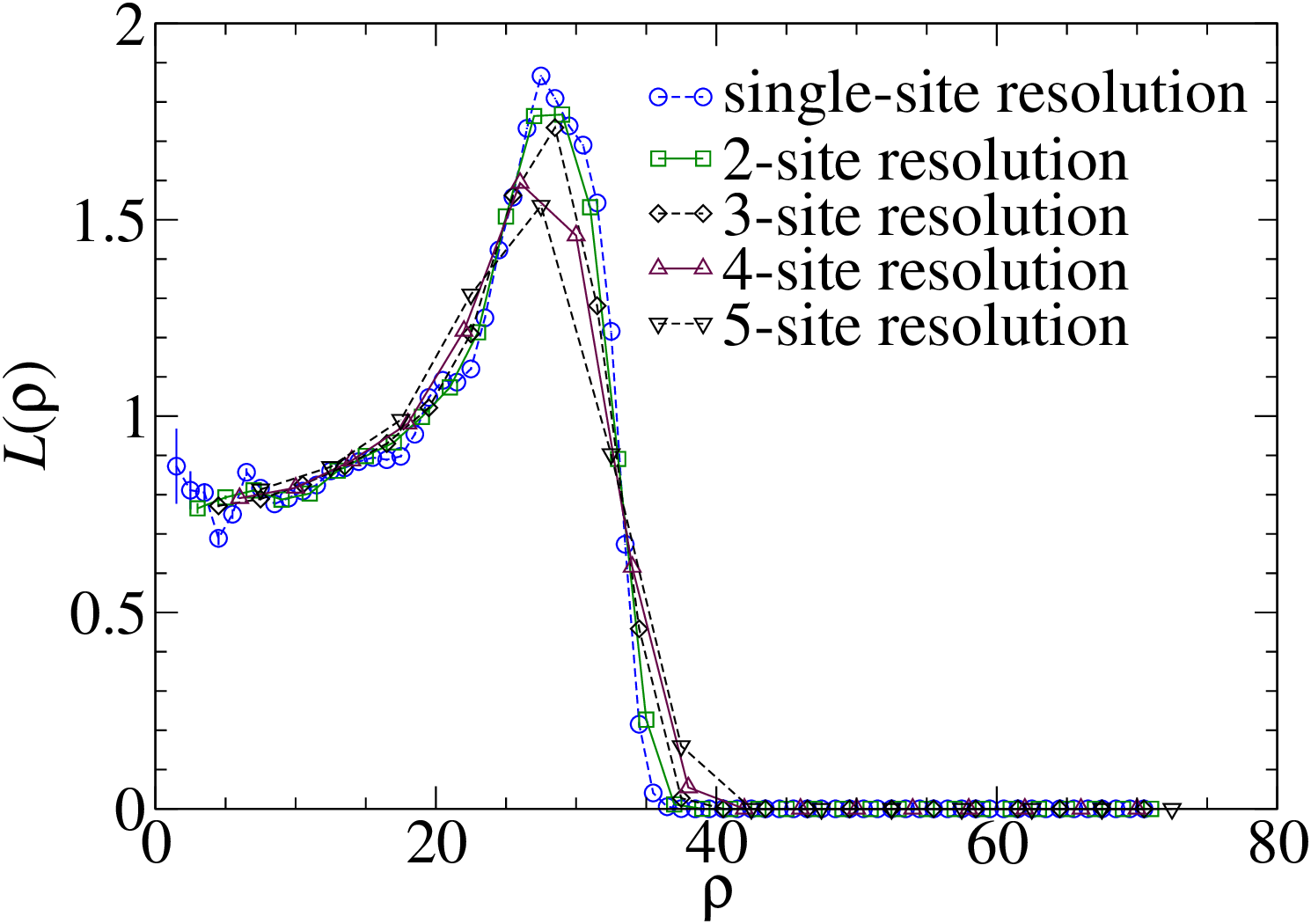}
\hfill
\includegraphics[width=8.5cm]{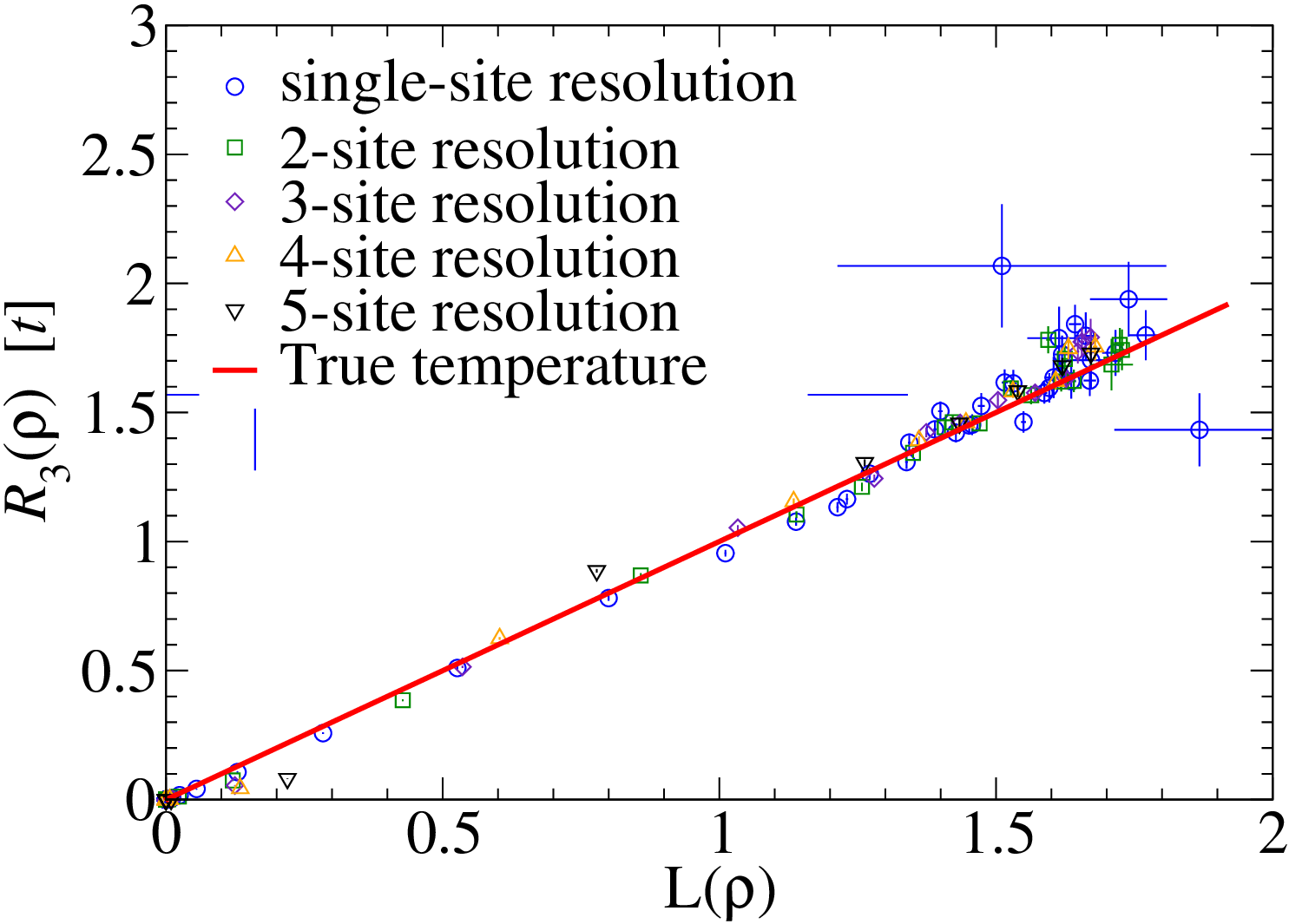}
\hfill
\includegraphics[width=8.5cm]{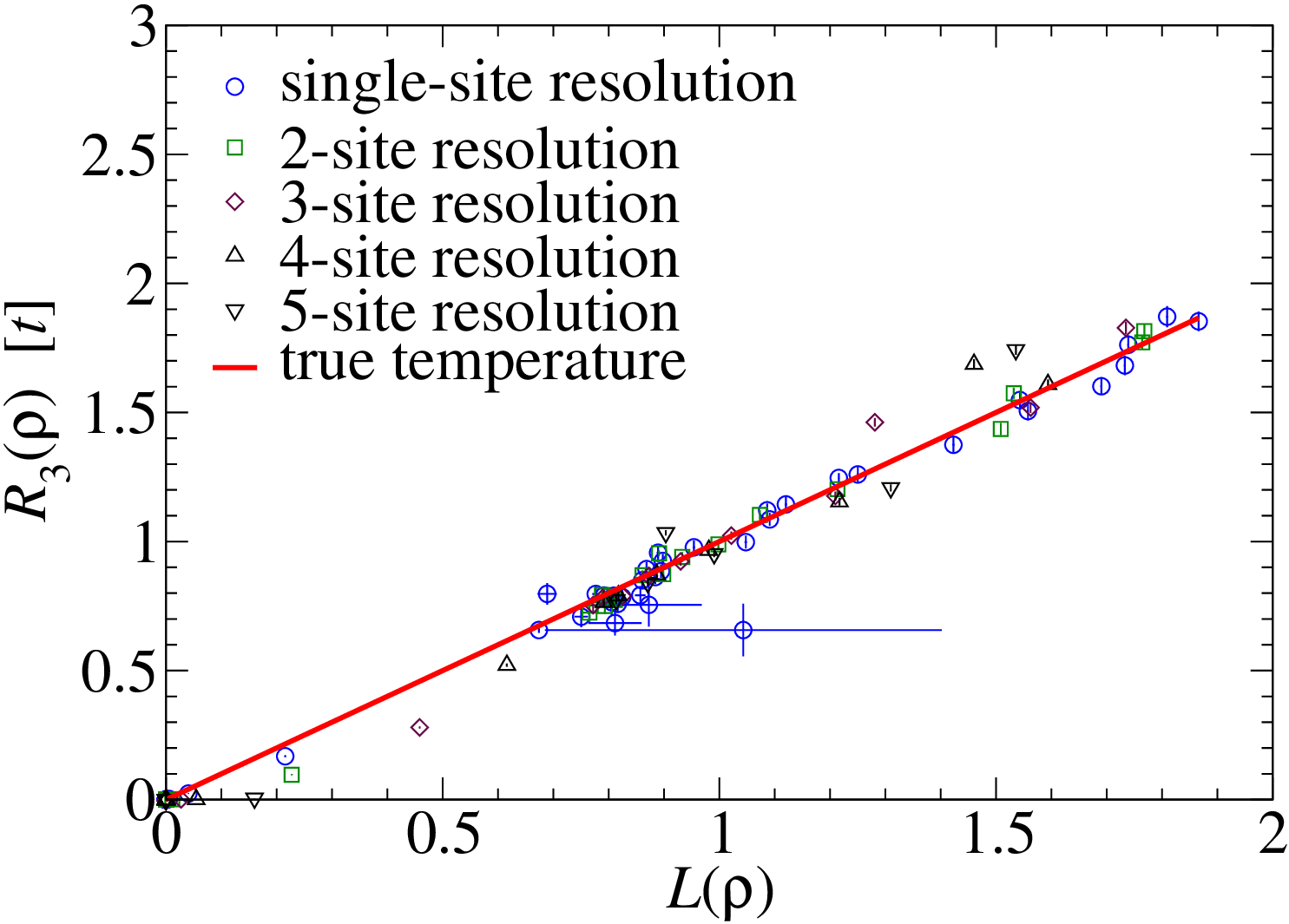}
\hfill
\caption{The fluctuation-dissipation thermometry scheme remains applicable to in-situ density experiments which have a resolution of a few sites.  The system consists of 3D optical lattice with $N=125,000$ bosonic ${}^{87}$Rb atoms. Increasing the binwidth from 1.0 to 5.0 in steps of 1.0 increases the systematic error in the dissipation term $L(\rho)$, but the temperature estimate remains realiable. Top panels show the dissipation term $L(\rho)$ and bottom panels show the $L(\rho)$ vs $R(\rho)$ curves. The parameters in the left column are $U/t = 10$, $T/t = 1$; and in the right column $U/t = 50$, $T/t = 1$, with a window size $\xi = 3$. We average over a 1000 independent measurements obtained from a QMC simulation. }
\label{fig:LR_binwidth}
\end{figure*}

Although we aim at single-site resolution detection tools in the analysis of the fluctuation-dissipation thermometry scheme in this paper, the scheme remains applicable when the resolution is just a few sites, as illustrated in Fig.~\ref{fig:LR_binwidth}.
Here, we simulate such experiments with different resolutions up to 5 sites, by varying the bin. The fluctuation dissipation thermometry scheme suffers from increasing systematic errors in the dissipation term $L(\rho)$ both for $U/t=10$ and $U/t=50$ at $T/t=1$. Yet, a relatively linear $L(\rho)-R(\rho)$ relationship could still be observed on average. For these cases alone, the estimated temperatures do not deviate more than 10\% as shown in Table \ref{table:resolution}.
\begin{table} [!htp]
\centering
\begin{tabular}{| c | c |}
\hline
n-site resolution  & estimated        \\
(binwidth)         & temperature [t]  \\
\hline\hline
\multicolumn{2}{|c|}{({\it U/t=10 , T/t=1}):} \\
\hline
1  &  0.977 $\pm$ 0.007  \\
2  &  0.990 $\pm$ 0.006  \\
3  &  0.997 $\pm$ 0.006  \\
4  &  1.016 $\pm$ 0.007  \\
5  &  1.048 $\pm$ 0.007  \\
\hline\hline
\multicolumn{2}{|c|}{({\it U/t=50 , T/t=1}):} \\
\hline
1  &  0.994 $\pm$ 0.008 \\
2  &  1.014 $\pm$ 0.007 \\
3  &  1.032 $\pm$ 0.007 \\
4  &  1.045 $\pm$ 0.008 \\
5  &  1.094 $\pm$ 0.008 \\
\hline
\end{tabular}
\caption
{
The estimated temperatures obtained in simulations for various resolutions.  We take a 3D optical lattice system with $N=125,000$ ${}^{87}$Rb atoms and a window size $\xi = 3$. We average over 1000 independent measurements obtained from a  QMC simulation. In the top panel, $ U/t = 10$ , $T/t = 1$; in the bottom panel, $ U/t = 50$, $T/t = 1$. The estimated temperatures do not deviate more than 10\% despite the systematic error involved.
}
\label{table:resolution}
\end{table}
However, when the resolution is worse than 5 sites uncontrolled systematic errors dominate and the scheme fails.

\section{Fluctuations in temperature $T$ and total particle number $N$}
\label{appendix_fluctuations}

\begin{figure}[!htp]
\centering
\includegraphics[width=7.5cm]{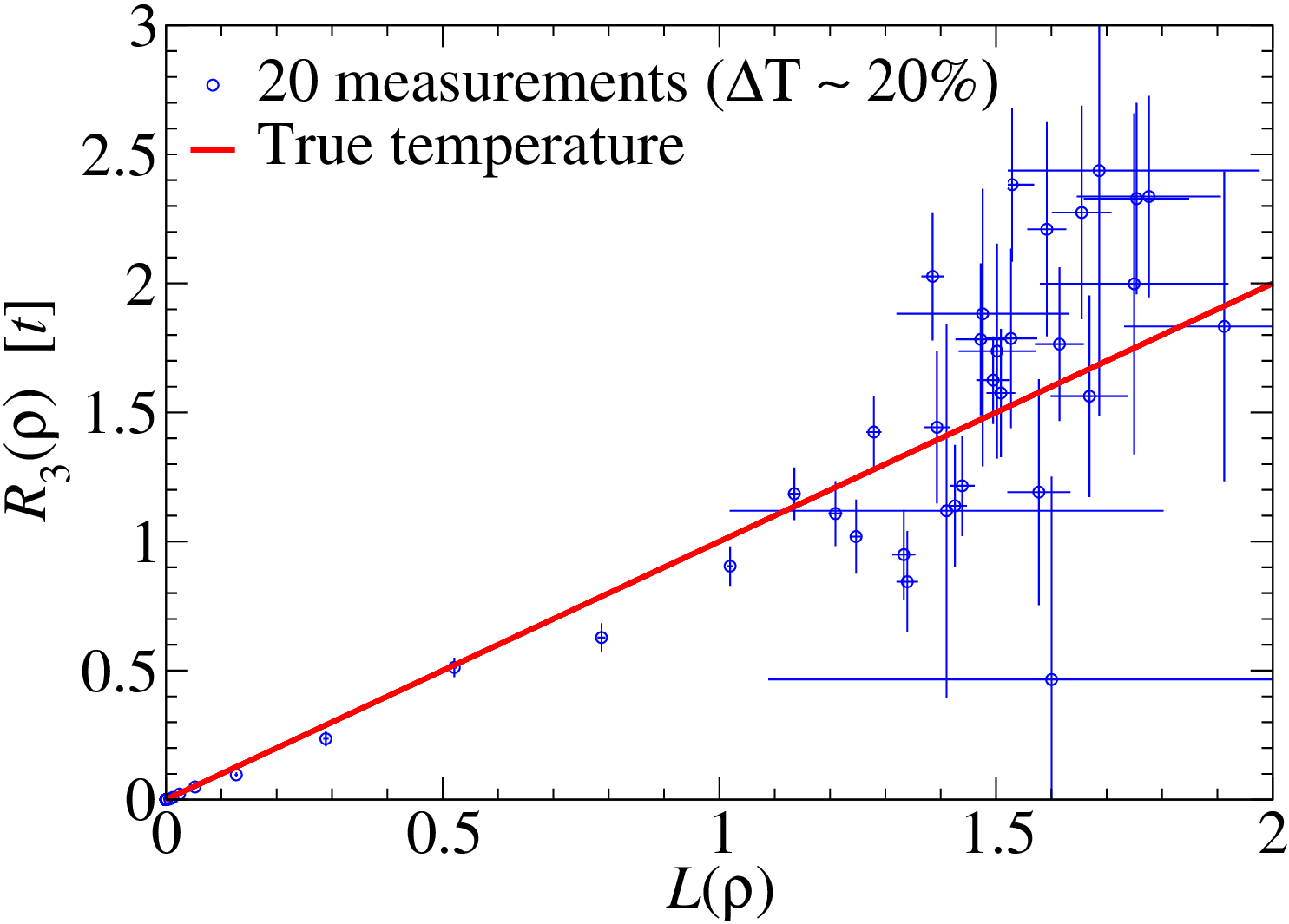}
\caption
{
The fluctuation dissipation thermometry scheme can still give a reliable estimate for the temperature when it has a spread $\Delta T / T \sim 20$\%.
The parameters for the  3D bosonic ${}^{87}$Rb optical lattice system are $N = 125,000$, $U/t = 10$, $T/t = 1$, $\xi = 3$, and we took 20 independent measurements from a QMC simulation. 
}
\label{fig:vary-T}
\end{figure}
\begin{figure}[!htp]
\centering
\includegraphics[width=7.5cm]{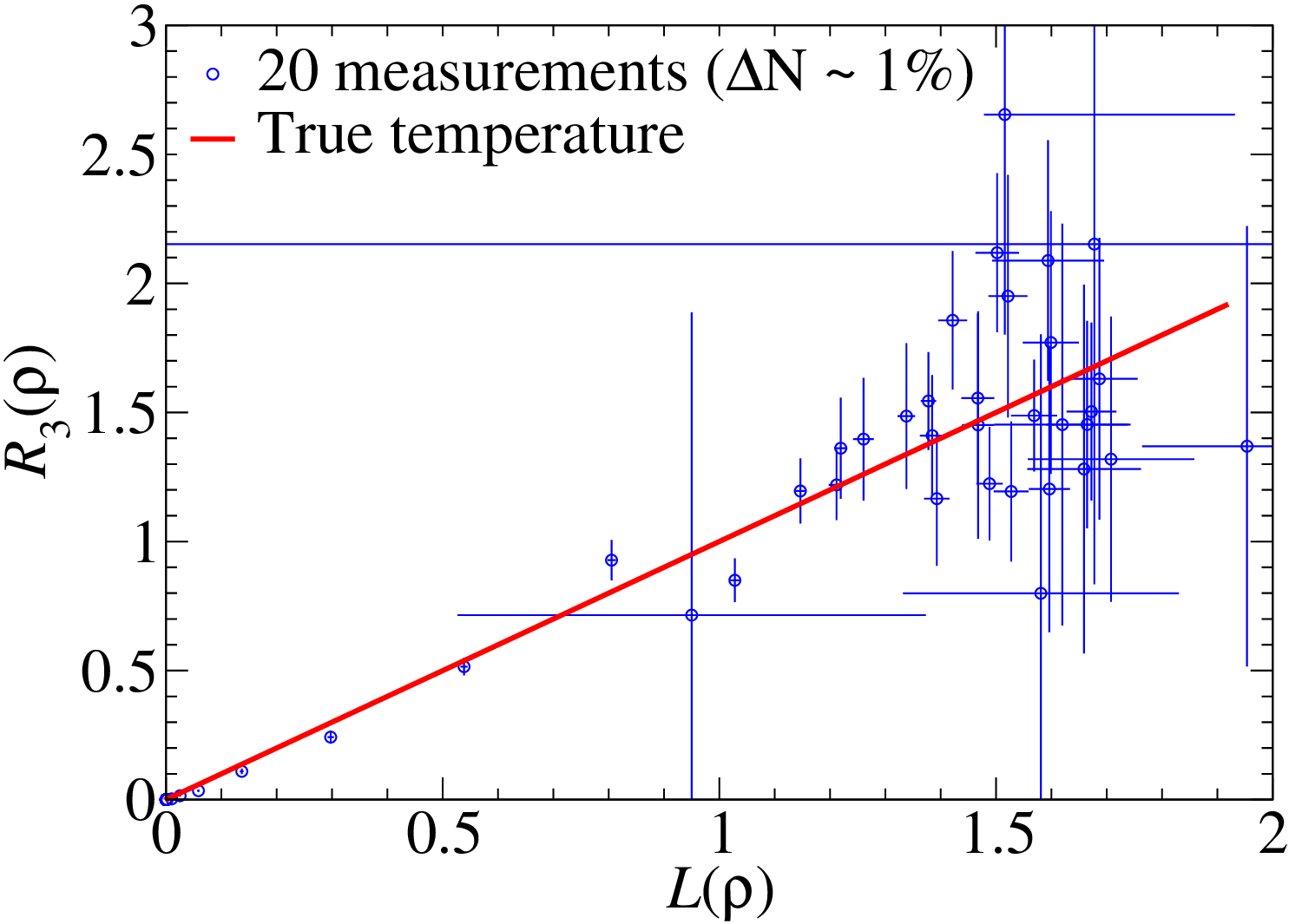}
\caption
{
The fluctuation-dissipation thermometry scheme remains valid for a 3D bosonic ${}^{87}$Rb system with total particle number $N = 125,000$ and spread $\Delta N$ of about 2,000. The parameters are $U/t = 10$, $T/t = 1$, $\xi = 3$,  and we took 20 independent measurements from a QMC simulation.
}
\label{fig:vary-mu}
\end{figure}
Different runs of an experiment will have small temperature and particle number changes. We model this by averaging over simuations where the temperature $T$ and the total particle number $N$ fluctuate. First, by letting $T$ fluctuate within a spread $\Delta T/T$ of about 20\%, we observe no qualititive difference in the thermometry scheme, as illustrated in Fig.~\ref{fig:vary-T} for the case of $U/t=10$, $T/t=1$ where we find an estimated temperature $T/t= 0.973(47)$.

Second, we let the chemical potential $\mu$ fluctuate such that the total particle number $N$ fluctuates with a spread $\Delta N$. We observe that a spread up to 1-2\% for the case of $U/t=10, T/t=1, N=125,000$ still yields a quantitatively reliable estimate for the temperature, $T/t=1.040(26)$. However, this scheme will have systematic errors for fluctuations bigger than 3\%.